\documentclass[graybox,openany]{svmult}

\usepackage{mathptmx}       
\usepackage{helvet}         
\usepackage{courier}        
\usepackage{type1cm}        
\usepackage{makeidx}         
\usepackage{graphicx}        
\usepackage{multicol}        
\usepackage[bottom]{footmisc}
\usepackage{color,colordvi}
\makeindex             

\newcommand\ebv{\rm E(B-V)}
\newcommand\HI{\rm HI}
\newcommand\HeI{\rm HeI}

\newcommand\microG{$\mu$G}
\newcommand\NHI{\rm N(HI)}
\newcommand\NH{\rm N(H)}

\newcommand\np{$\mathrm{n_{p}}$}
\newcommand\nHI{$\mathrm{n_{HI}}$}

\newcommand\MgII{\hbox{Mg$^+$}}
\newcommand\FeII{\hbox{Fe$^+$}}

\newcommand\nel{n(e)}
\newcommand\cc{cm$^{-3}$}

\newcommand\msun{\ensuremath{\rm M_{\odot}}}

\newcommand\be {\begin{equation}}
\newcommand\en{\end{equation}}

\newcommand\carbtwelve{\ensuremath{\rm ^{12}C\ }}
\newcommand\oxysixteen{\ensuremath{\rm ^{16}O\ }}
\newcommand\cfourteen{\ensuremath{\rm ^{14}C\ }}
\newcommand\fesixty{\ensuremath{\rm ^{60}Fe\ }}
\newcommand\fetsix{\ensuremath{\rm _{26}Fe\ }}
\newcommand\fefeight{\ensuremath{\rm ^{58}Fe\ }}
\newcommand\fefsix{\ensuremath{\rm ^{56}Fe\ }}

\newcommand\neontwentytwo{\ensuremath{\rm ^{22}Ne\ }}
\newcommand\neontwenty{\ensuremath{\rm ^{20}Ne\ }}
\newcommand\altwentysix{\ensuremath{\rm ^{26}Al }}
\newcommand\ziforty{\ensuremath{\rm _{40}Zi\ }}
\newcommand\nickelfn{\ensuremath{\rm ^{59}Ni\ }}
\newcommand\cobaltfn{\ensuremath{\rm ^{59}Co\ }}
\newcommand{\commentpcf}[1]{}
\newcommand{\delete}[1]{}
\newcommand{\move}[1]{}
\newcommand {\gtsim} {\ {\raise-.5ex\hbox{$\buildrel>\over\sim$}}\ }
\newcommand {\ltsim} {\ {\raise-.5ex\hbox{$\buildrel<\over\sim$}}\ }

\newcommand\kms{\hbox{km s$^{-1}$}}

\newcommand\cmtwo{\hbox{cm$^{-2}$}}
\newcommand{\aj}{{AJ}}
\newcommand{\araa}{{Ann.\ Rev.\ Astron.\ Astrophys.}}
\newcommand{\apj}{{ApJ}}
\newcommand{\apjl}{{ApJ Lett.}}
\newcommand{\apjs}{{ApJS}}
\newcommand{\apss}{{Ap\&SS}}
\newcommand{\aap}{{A\&A.}}

\newcommand{\mnras}{{MNRAS}}

\newcommand{\ssr}{{Space~Sci.~Rev.}}

\newcommand{\nat}{{Nature}}

\newcommand{\jgr}{{J.~Geophys.~Res.}}

\begin{document}
\title*{Effect of Supernovae on the Local Interstellar Material}
\titlerunning{Supernovae and local ISM }
\author{Priscilla Frisch and Vikram V.~Dwarkadas }
\institute{Priscilla Frisch \at University of Chicago, Department of Astronomy and Astrophysics, 5640 S Ellis Ave, Chicago, IL 60637, \email{frisch@oddjob.uchicago.edu}
\and Vikram V.~Dwarkadas \at University of Chicago, Department of Astronomy and Astrophysics, 5640 S Ellis Ave, Chicago, IL 60637, \email{vikram@oddjob.uchicago.edu}}
 
%
\maketitle
\abstract{ A range of astronomical data indicates that ancient
  supernovae created the galactic environment of the Sun and sculpted
  the physical properties of the interstellar medium near the
  heliosphere. In this paper we review the characteristics of the
  local interstellar medium that have been affected by supernovae.
  The kinematics, magnetic field, elemental abundances, and
  configuration of the nearest interstellar material support the view
  that the Sun is at the edge of the Loop I superbubble, which has
  merged into the low density Local Bubble. The energy source for the
  higher temperature X-ray emitting plasma pervading the Local Bubble
  is uncertain.  Winds from massive stars and nearby supernovae,
  perhaps from the Sco-Cen Association, may have contributed
  radioisotopes found in the geologic record and galactic cosmic ray
  population.  Nested supernova shells in the Orion and Sco-Cen
  regions suggest spatially distinct sites of episodic star formation.
  The heliosphere properties vary with the pressure of the surrounding
  interstellar cloud. A nearby supernova would modify this pressure
    equilibrium and thereby severely disrupt the heliosphere as well
  as the local interstellar medium.  }

\pagestyle{myheadings}
\tableofcontents
\markboth{}{ }
\section{Introduction}
Nearly a century ago Harlow Shapley \cite{Shapley:1921} noticed that
the Sun is traveling away from Orion and speculated that variations in
the brightness of Orion stars could result from encounters between the
stars and nebulosity that would ``gravely affect the atmosphere
surrounding any attendant planet''.  By analogy he suggested that were
the Sun to encounter diffuse nebulosity it could induce severe changes
in the terrestrial climate.  Such an idea is not far-fetched.
Supernovae and winds from massive stars inject large amounts of energy
into the interstellar medium and reshape and remix interstellar
material with {shock waves} and expanding {bubbles}. Nearby {supernovae}
occur at rates that are larger than the galactic average due to the
solar location inside of {Gould's Belt}.  The flux of galactic {cosmic
rays} into the terrestrial atmosphere depends on the response of the
{heliosphere} to the ambient {interstellar medium}
\cite{FlorinskiZank:2006jos,MuellerFrisch:2006apj}.  Geologic records
indicate that cosmic rays from supernova have penetrated to the
surface of the Earth {\cite{Brakenridge:1981}.  The Sun is traveling
  toward the constellation of Hercules at a velocity of $\sim 18$ pc
  per Myr through the {local standard of rest} ({LSR})
  \cite{Frisch:2015ismf3}.  Without question the supernovae that
  generated the {Loop I} {superbubble} \cite{Berkhuijsen_etal_1971}
  affected the ambient interstellar medium and possibly the
  interplanetary environment of the Earth. Locations of many
  supernovae within 400 pc and over the past $\sim 2$ Myr have been
  identified.

The Sun is immersed in a small ($<15$ pc) cluster of low density
partially ionized interstellar cloudlets of the type that were once
identified as the ``{intercloud medium}'' because of the low
extinctions, \ebv$< 0.001$ mag.  Most local gas is warm, $ T = 5000 -
12,500$ K, low density partially ionized gas with log \NHI$< 18.7$
\cmtwo, $<$\nHI$> =0.01 - 0.10 $ \cc, and \np$\sim 0.1$
\cc\ (e.g. \cite{RogersonYork:1973ions,RLIIItemp,Frisch:2011araa}).  A
magnetic field with strength $\sim 3$ \microG\ shapes the heliosphere
\cite{SlavinFrisch:2008,Schwadron:2011sep}.  The only known local
cloudlet that is cold and neutral is the filamentary Local Leo Cold
Cloud (LLCC) dust cloud at distance $\sim 18$ pc
\cite{MeyerPeek:2012leo,Peeketal:2011}.

This review discusses the configuration of massive stars in Gould's
Belt that spawn nearby supernovae (Section \ref{sec:gb}), bubble
formation (Section \ref{sec:bubbles}), the location of the Sun inside a
superbubble rim (Section \ref{sec:loopI}) that merges into the low density
Local Bubble cavity (Section \ref{sec:lb}), the Orion superbubble (Section
\ref{sec:ori}), short-lived radioisotope clocks of recent nearby
supernova found in geological and astrophysical data (Section
\ref{sec:iso}), and the impact of supernovae on the heliosphere (Section
\ref{sec:helio}).

The Local Bubble is characterized by a cavity in the interstellar
medium (ISM).  Figure \ref{fig:lb} shows the distribution of interstellar
material within $\sim 400$ pc according to the cumulative reddening of
starlight as measured by the color excess \ebv\ (Frisch et al.~2015,\cite{Frisch:2015ismf3}, in preparation).
Superimposed on the reddening maps are 
nearby superbubble shells, the interarm interstellar magnetic field
(ISMF) indicated by pulsar data \cite{Salvati:2010}, and the local ISMF
direction diagnosed by the {Ribbon} of energetic neutral atoms (ENAs)
discovered by the Interstellar Boundary Explorer (IBEX) spacecraft
\cite{McComas:2009sci,Schwadron:2009sci,Funsten:2013,Heerikhuisen_etal_2010,Zirnstein:2016ismf}.
The extinction void around the Sun, denoted the Local Bubble, occurs where the interior of
Gould's Belt blends into the interarm regions of the third galactic quadrant.

\begin{figure}[th!]
\begin{center}
\includegraphics[scale=0.38]{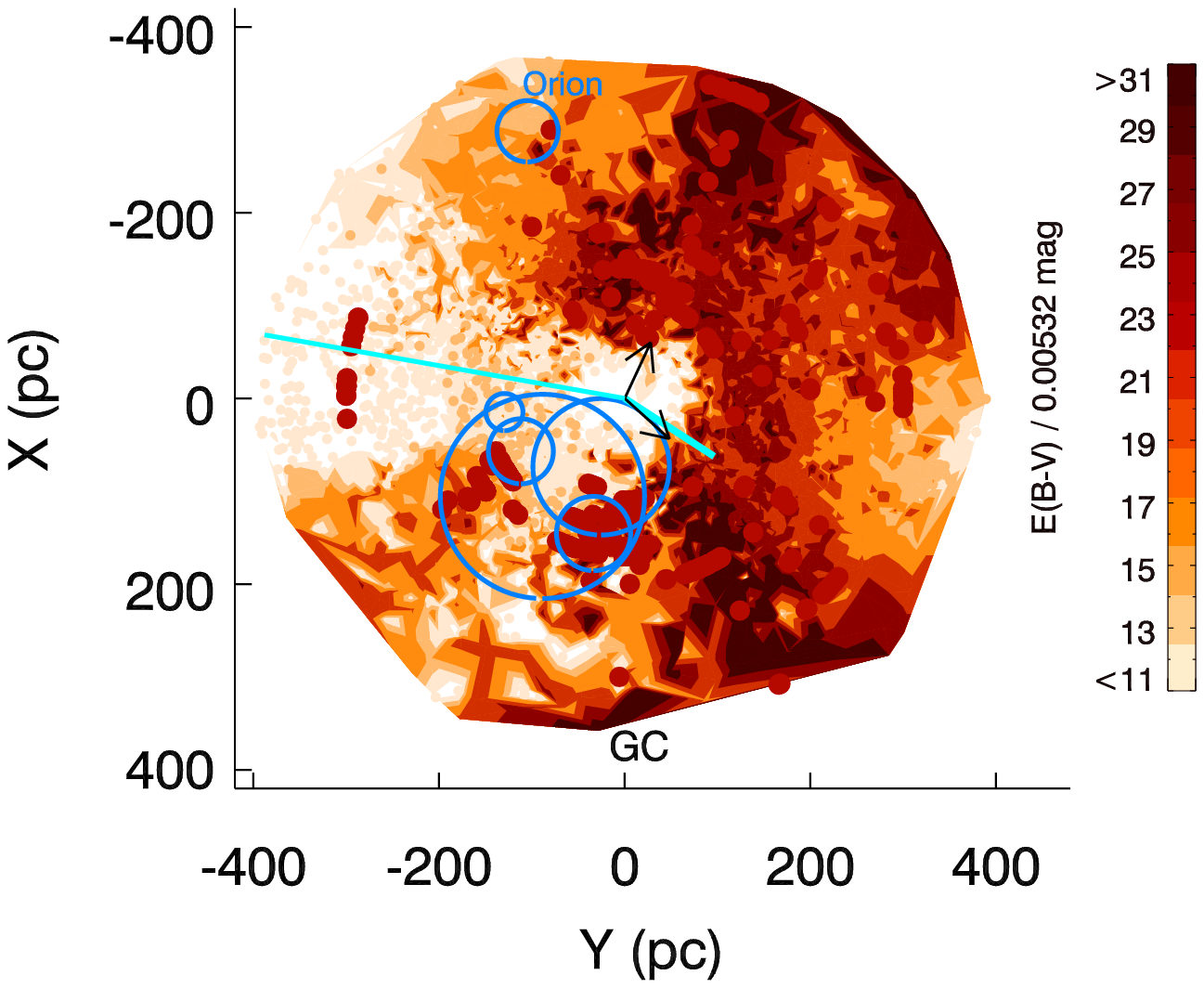}
\includegraphics[scale=0.38]{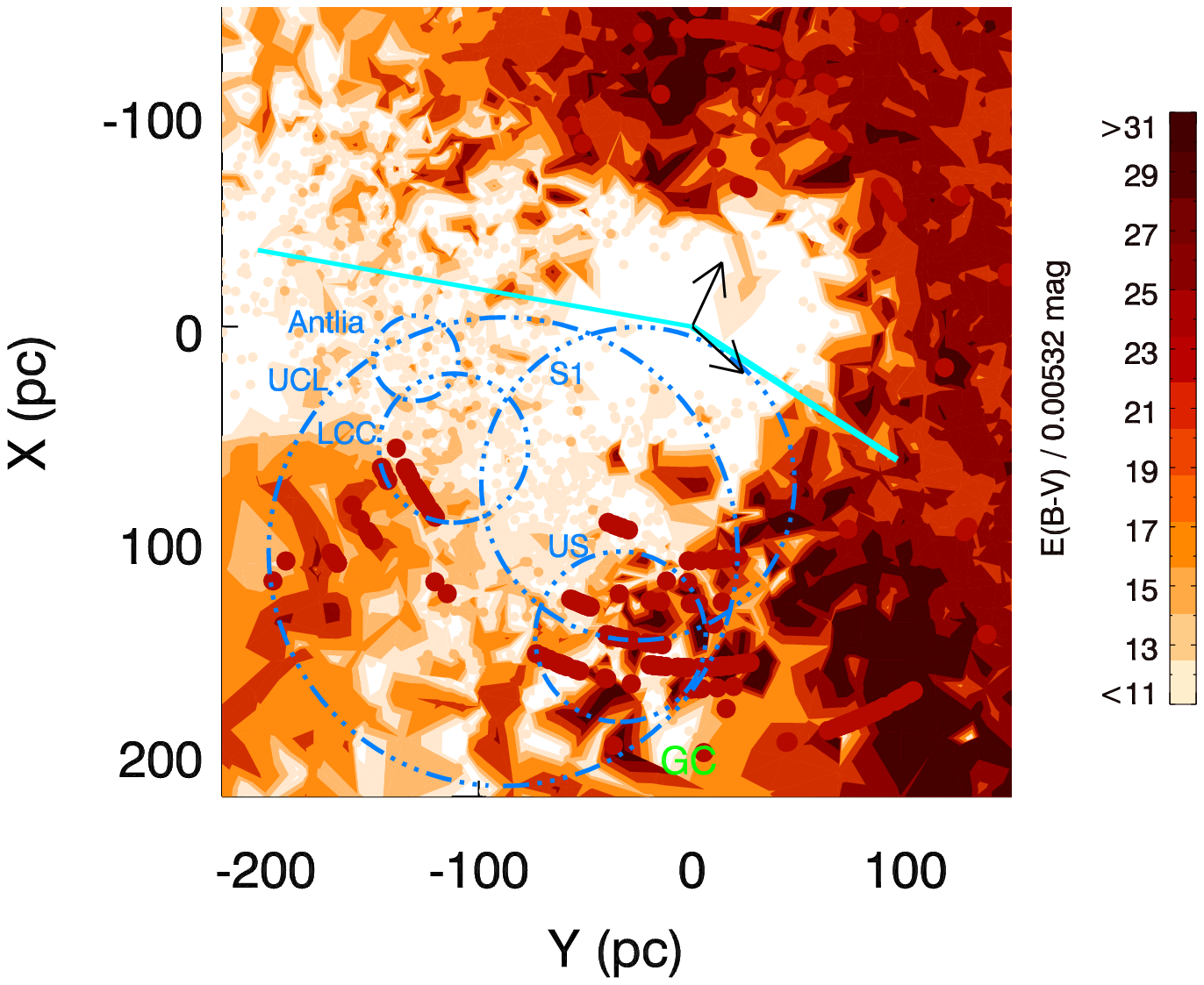}
\end{center}
\caption{{The Local Bubble:} The cumulative extinction of interstellar
  dust is shown over spatial scales of $\sim 300$ pc (right) and $\sim
  800$ pc (left) projected onto the galactic plane (from \cite{Frisch:inprep}).  
The galactic center is at large positive
  values of X and the galaxy rotates toward positive Y values.  The
  cumulative extinction is determined from the color excess
  \ebv\ measured for each star where the astrometric distances are
  required to agree with distances obtained from \ebv\ and the
  spectral type of the star (see Frisch et al., 2015, \cite{Frisch:2015ismf3}
  for more information).  The color bar shows color excess \ebv\
in units of 0.00532 mag.  
  Round dark maroon circles show the locations of nearby
  molecular {CO} clouds.  The two black arrows show the {LSR} velocities
  of the {CLIC} (perpendicular to the {S1 shell}) and the Sun (roughly
  tangential to the S1 shell).  The long and short cyan-colored lines
show the directions of the interarm magnetic field and {IBEX} Ribbon magnetic
field directions, respectively
  \cite{Funsten:2013,Frisch:2015ismf3}.  The circles show 
the three superbubble shells in Sco OB2, the Antlia SN remnant, and  the Ori-Eri
  superbubble (left figure only), and the S1 shell (see text for
  details).  
}
\label{fig:lb}
\end{figure}

\section{Gould's Belt and Massive Stars } \label{sec:gb}

Nearby clusters of O--B2.5 massive stars that are progenitors of
core-collapse Type II and Type Ib/c supernovae form a thin planar
ring-like structure around the Sun known as ``Gould's Belt''
\cite{PerrotGrenier:2003,Grenier:2004rev,Bobylev:2014gbrev}.  Gould's
Belt is part of a large-scale warp in the distribution of young stars
in the galactic plane \cite{Alfaro:1991}.  The traditional
configuration of Gould's Belt as an inclined plane defined partly by
the Sco OB2 and Orion OB1 associations is shown in Figure \ref{fig:gb}
(from Grenier 2004, \cite{Grenier:2004rev}).  Gould's belt is tilted by
an angle of $\sim 17.2^\circ$ with respect to the galactic plane, with
the ascending node toward $\sim 296^\circ$, and centered $\sim 104$ pc
away toward $\ell \sim 184^\circ$, where the uncertainties arise from
the different selection criteria for testing the Gould's Belt
configuration
\cite{StothersFrogel:1974,PerrotGrenier:2003,Grenier:2004rev,Bobylev:2014gbvdbeltorionarm}).
An alternate perspective compares the distribution and kinematics of
open clusters in the Orion OB1 association with those of Sco OB2 to
characterize {Orion OB1} as belonging to the {Local Arm} where high
densities of open clusters and ongoing star formation appear, in
contrast to the Sco OB2 stars that are located on the outskirts of the
Local Arm with lower densities of active star-forming regions
\cite{Alfaro_etal_2009,Elias_etal_2009}.  Elias et
al. \cite{Elias_etal_2009} establish that the Local Bubble region
around the Sun is devoid of open clusters in comparison to the Orion
region.  Bobylev and Bajkova (2014,
\cite{Bobylev:2014gbvdbeltorionarm}) used astrometric data to define
the Orion arm as a narrow ellipsoid directed toward $\ell = 77.1^\circ
\pm 2.9^\circ$ with a symmetry plane inclined to the Galactic plane by
$5.6^\circ \pm 0.2^\circ$, and with a longitude of the ascending node
of the plane toward $70^\circ \pm 3^\circ$
\cite{Bobylev:2014gbvdbeltorionarm}.  Regardless of the detailed
description of Gould's Belt versus the Local Arm, most of the
early-type O--B2.5 massive stars in the solar vicinity coincide with
the traditional configuration of Gould's Belt.

\begin{figure}[h!]
\begin{center}
\includegraphics[scale=.47]{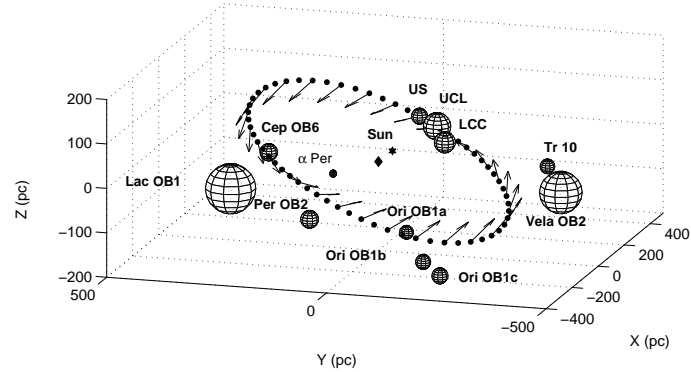}
\end{center}
\caption{{Gould's Belt: } Stellar associations of Gould's Belt and the
  velocity field of the stars with respect to the LSR
  are shown in 3D.  The diamond shows the center of Gould's Belt
  and the star shows the location of the Sun.  The galactic center is
  directed toward large positive values for the x-coordinate.  Figure
  from Grenier (2004, \cite{Grenier:2004rev}).}
\label{fig:gb}
\end{figure}

About 17--20 supernovae per million years formed in the entire Gould's
Belt during the past several million years, which is a rate 3--4 times
that of the local galactic average \cite{Grenier:2004rev}.  Since 80\%
of galactic supernova are from the core collapse of massive stars, the
SNe in Gould's Belt will account for the most likely supernova to have
shaped the physical properties of the local interstellar medium.  Over
tens of {millions of} years these supernovae have rearranged the
interstellar material near the Sun into the networks of filaments,
arcs, shells and superbubbles that are observed
\cite{Heiles:1979,Heiles:1984worm,cox05}.

\section{Supernova Remix of the Interstellar Medium}  \label{lab:remix}

Superbubble shells, either complete, or incomplete ``worm'' or
filamentary structures, are common in the ISM and are nearly all
explained by energy injection from stellar winds and supernovae
(Heiles 1984, \cite{Heiles:1984worm}).  Measurements of the Zeeman
splitting of the \HI\ 21-cm line shows that morphologically distinct
filaments generally consist of warm neutral or partially ionized gas
where magnetic pressure dominates thermal pressures by a factor of
$\sim 67$, and turbulent pressure by a factor of $\sim 10$
\cite{Heiles:1989shellbeta}.  The rapid evolution of massive stars in
Gould's Belt has frequently altered the ISM in the solar neighborhood.

\subsection{Bubbles and Superbubbles} \label{sec:bubbles}
Supersonic winds from massive stars can evacuate large regions of
space around the star, with radii of several parsecs. The detailed
structure of these `wind-blown bubbles' was first elucidated by Castor
et al.~ (1975, \cite{castoretal75}) and Weaver et al.~ (1977,
\cite{weaveretal77}), and has been subsequently refined and discussed
by many authors \cite{mvl84, km92a, km92b}.  In general, the bubbles
consist of a very low density ($<$ 0.01 particles cm$^{-3}$ on
average) interior surrounded by a dense {shell} of material, bounded by
a radiative shock that serves as the boundary of the bubble. The
bubbles can be either density or ionization bounded - if the shell is
dense enough, an ionization front will be trapped in the dense shell,
beyond which neutral material can be found.  The ISM magnetic field
can also affect the size, shape and evolution of circumstellar bubbles
\cite{vanmarleetal15}. Weaver et al. (1977, \cite{weaveretal77})
showed that the radius of the bubbles primarily grows with time as
$R_b \propto \left[\frac{L}{\rho}\right]^{1/5} t^{3/5}$ where $L=0.5
\dot{M}v_w^2$ is the mechanical luminosity of the wind with mass-loss
rate $\dot{M}$ and wind velocity $v_w$, and $\rho$ is the density of
the ambient medium. 

Numerical simulations have been successful in
confirming the analytical predictions and reproducing the general
structure, formation and morphology of massive-star bubbles
\cite{gml96a, glm96b,fhy03,fhy06,vvd07}. Other simulations have
explored the evolution of the subsequent supernova shock waves within
the bubbles
\cite{tbfr90,trfb91,rtfb93,vvd05,vvd07,vanmarleetal10}. Although the
highly supersonic winds around massive stars (wind velocity on order
1000-2000 km s$^{-1}$), and the low density and high pressure
pervading the bubble would point to an extremely high temperature
within the bulk of the bubble (of order 10$^7$ to 10$^8$ K), X-ray
observations have shown that if hot gas is detected at all, its
temperature and emission measure are both low, on the order of a few
times 10$^6$ K~\cite{cgg03, chuetal03, toalaetal15, toalaetal16}. Many
authors have tried to simulate these observations to understand the
origin of the low temperatures, with some degree of success in
matching the bubble temperatures and their X-ray spectra
\cite{ta11,dr13,dr15}.

Clusters of massive stars group together to form an association.
The correlated supernovae resulting from the explosion of these stars
can form even larger bubbles due to the combined effects of the winds
and supernova explosions. The aptly named superbubbles \cite{mm88,
  mm89, yadavetal16} rearrange the morphology and physical
characteristics of the surrounding ISM.  Heiles (1979,
\cite{Heiles:1979}) defines bubbles with injected energies greater
than $3 \times 10^{52}$ ergs as {supershells} or {superbubbles}.  The
structure of these superbubbles is similar to their smaller brethren,
and can be approximated from the bubble theory allowing for continuous
energy input from an association of stars and their resulting SNe.
MacLow and McCray (1988, \cite{mm88}) have shown that the bubble
approximation is valid even for superbubbles, and show that the
superbubble radius can be written as $R_{sb} \sim 267
\left[\frac{L_{38}t_7^3}{n_0}\right]^{1/5}$ pc, where $L_{38}$ is the
mechanical luminosity in terms of $10^{38}$ ergs s$^{-1}$, $t_7$ is
the time in units of 10$^7$ years, and $n_0$ is the atomic number
density. Due to their extremely large size, it is clear that these
bubbles are not expanding in a homogeneous interstellar medium but in a
medium whose structure is constantly being stirred due to heating by
supernova explosions \cite{korpietal99}.

The expanding flow sweeps up interstellar material and magnetic fields
into a postshock shell. The mass within the bubble interior is likely
regulated by evaporation from the cool bubble walls, by entrainment
and ablation from denser clouds remaining within the bubble, and by
the penetration of ambient interstellar clouds that are not destroyed
by the photoevaporative effect of the massive star
\cite{shull93,og04}.  A superbubble shell thickens as it sweeps up
magnetic field lines during the pressure-driven snowplow stage,
producing regions in the evolved shell where the ISMF is perpendicular
to the gas velocity
\cite{FerriereMacLow:1991,HanayamaTomiska:2006,maclow99}, such as
is found for the immediate solar environment (Section \ref{sec:loopI}).
For a cylindrical model with the ISMF parallel to the axis of the cylinder,
superbubble expansion parallel to the radial direction produces a
configuration where gas velocities are perpendicular to the ISMF
direction \cite{HanayamaTomiska:2006}.

It is in the context of these known bubbles that superbubbles in our
vicinity, and the detection of a low density, high temperature region
around our solar system, the Local Bubble, need to be evaluated.

\subsection{Radio Superbubbles and Magnetic Loops }\label{ref:loopI}
The closest superbubbles that have influenced the local ISM occurred
in the Sco-Cen association, and were first revealed through
{observations of} the intense radio continuum source known as the
{North Polar Spur} (NPS) that extends north from longitude $\ell \sim
30^\circ$.  The NPS was part of a loop-like structure with a
non-thermal continuum, that was modeled as a supernova remnant likely
to be within 100 pc \cite{BrownDaviesHazard:1960,Davies:1964}.
Berkhuijsen et al. (1971,
\cite{Berkhuijsen:1971,Berkhuijsen_etal_1971}) identified four
non-thermal radio loops at 830 MHz, Loops I--IV, with the NPS the
brightest part of Loop I (see early review by Salter,1983,
\cite{Salter:1983}).  The strongly polarized radio Loop I {(64\% to
  72\% polarization)}, indicates a magnetic field that is uniform in
direction \cite{Heiles_etal_1980nps}.  Loops II and III are
radio-continuum features \cite{Berkhuijsen:1971,Kun:2007loopIII} while
Loop IV coincides with the extended HII region around the nearby B1V
star Spica indicating that Loop IV does not have a supernova origin
\cite{Reynolds:1984alpvir}.  Using skymaps from the Wilkinson
Microwave Anisotropy Probe (WMAP) at 23, 33, and 41 GHz, Vidal et
al. \cite{VidalDickinsonDaviesLeahy:2015} identified the original four
loops and eleven new non-thermal loop structures, some prominent only
through polarization.  The mean spectral index for the brightness
temperature {of} the polarized emissions is $-3.06 \pm 0.02$,
verifying the synchrotron nature of the emission from these radio
loops.

\begin{figure}[h!]
\begin{center}
\hspace*{-0.1in}
\includegraphics[scale=.65]{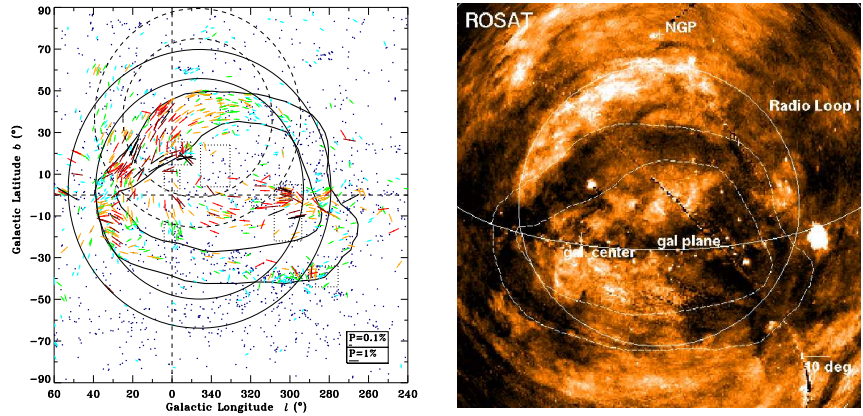}
\end{center}
\caption{ {Left:  Polarized starlight toward Loop I:} 
The optical polarizations that trace the
ISMF in the neutral gas that shadows the soft X-ray background
(right) are shown together with the
S1/S2 superbubble shells (solid/dashed circles respectively) from Wolleben (2007, \cite{Wolleben:2007}).
The irregular lines reproduce the boundaries of the ``interaction ring'' shown at right.
Figure from Santos et al. 2011, \cite{Santos:2011}.
{Right:  Egger interaction ring:} 
Soft X-ray shadow appearing in ROSAT data in the energy range 0.1--2.0
keV.  The X-ray shadow is caused by neutral gas with a density of $\sim
15$ \cc, and aligned dust grains that polarized background starlight.
between two interstellar bubbles. 
Also identified are the location of Loop I (solid white line), the
interaction ring between the Local Bubble plasma and the Loop I bubble
(outlined by dashed lines), and galactic landmarks.  
The X-ray shadow corresponds to column densities that
jump from \NH$\sim 10^{20}$ \cmtwo\ to $7 \times 10^{20}$ \cmtwo.
Optical and UV data indicate a
neutral wall with \NH$\sim 10^{20}$ \cmtwo\ at a distance of $40 \pm
25$ pc.  Figure from Egger and Aschenbach 1995, \cite{Egger:1995}, and courtesy of the Max-Planck-Institut fur extraterrestrische Physik \url{(http://www/mpi/mpg/de/)}.}
\label{fig:santos}
\end{figure}

Filling in the picture of the influence of supernova on the local ISM
requires data on the neutral ISM.  Heiles \cite{Heiles:1979}
identified sixteen stationary \HI\ shells within 500 pc that have
diameters $\le 36^\circ$.  Eleven large interstellar shells beyond 500
pc were termed ``supershells''.  Radio-recombination lines showed that warm, 7000 K,
partially ionized filamentary structures are common throughout the ISM
\cite{HeilesReachKoo:1996}.  }

The largest diameter radio bubble in the sky is Loop I.  Successive
epochs of star formation in the parent molecular cloud of Sco OB2
created large-scale nested interstellar \HI\ shells that have provided
the basis for linking nearby bubble-like structures, such as Loop I,
with the parent clusters of stars \cite{deGeus:1991,Crawford:1991}.
Polarization data suggest these nested shells are closer to the Sun,
{(at a distance $\sim 100$ pc)}, in the region $20^\circ < \ell <
40^\circ$ than in the region $290^\circ < \ell < 310^\circ$ where
distances are $> 200$ pc \cite{Santos:2011}.  The three subgroups of
the Scorpius-Centaurus Association \cite{Blaauw:1964araa} are the
Upper Centaurus-Lupus (UCL), Lower Centaurus Crux (LCC), and Upper
Scorpius (US). Their nuclear ages were thought to be 14--15 Myr,
11--12 Myr and 4--5 Myr respectively (deGeus 1992, \cite{deGeus:1992},
see Figure \ref{fig:lb}). However, recent re-examination of the
evolutionary state and isochronal ages data by \cite{Pecaut2012} and
\cite{Pecaut2016} suggests that the subgroups are not consistent with
being simple, coeval populations which formed in single bursts, but
likely represents a multitude of smaller star formation episodes of
hundreds to tens of stars each. They have also re-evaluated the ages
and find them to be higher, at 16 Myr (UCL), 17 Myr (LCC) and 11 $\pm
1 \pm 2$ Myr (statistical, systematic) for US. The US age is twice as
large as previously assumed.  When stellar proper motions are
included, it is seen that the shell-forming events did not occur at
the present locations of the stellar subgroups, and the LCC is the
most likely source of the large-scale Loop I feature
\cite{Maiz-Apellaniz_2001}.  Frisch (1981, \cite{frisch81}) pointed
out that the low density interarm-type material near the Sun would
have led to asymmetric expansion of Loop I.  The X-ray remnant toward
the NPS would have resulted from star formation triggered by the
impact of a shock wave on the Aquila Rift dark cloud
\cite{Crawford:1991,deGeus:1991,deGeus:1992}.  Iwan
\cite{Iwan:1980npsXrayloopI} found that a reheated supernova remnant
was required to simultaneously explain the Loop I \HI\ radio shell and
the ridge of X-ray plasma, although she could not incorporate the
then-unknown foreground contamination of the X-ray background by solar
wind charge-exchange with interstellar neutrals
\cite{Koutroumpa:2009lowkeV}.

\subsection{The Local Bubble} \label{sec:lb}
The discovery of the soft X-ray background (SXRB, \cite{Bowyeretal:1968}) motivated
measurements of the X-ray spectra at low energies, $<2.5$ keV,
where a flat X-ray spectrum was found 
that limited the amount of possible interstellar absorption of the
X-ray photons \cite{McCammon:1983}.  The resulting ``displacement model''
required the X-rays to be produced inside a cavity in the neutral gas
\cite{williamsonetal74,sandersetal77}).  The
original interpretation of the {SXRB} data as tracing an evolved
supernova remnant has been reviewed in \cite{cr87,ms90}.  Interpreting
the physics of the hot gas has been surprisingly difficult because the
local source of the hot plasma could not be identified.  The supernova
explosion that produced the {Geminga} pulsar was initially suggested to
account for the soft X-ray emission \cite{Gehrels:1993}, but Geminga
was shown to originate near Orion instead
\cite{Frisch:1993,SmithCunha:1994,Pellizza:2005}.  It is now known
that the low energy X-ray spectrum is contaminated by foreground
emission from charge-exchange between {solar wind} plasma and
interstellar neutrals \cite{Koutroumpa:2009lowkeV}.

It had been suggested by Cox and Smith (1974, \cite{cs74}) that
supernovae could form and maintain a mesh of interconnected tunnels of
low density high temperature gas in the interstellar medium, producing
structures similar to the Local Bubble. Frisch (1981, \cite{frisch81})
argued that the data indicated that the local interstellar medium had
been processed by a shock front at least 2 Myr ago based on age limits
set by the soft X-ray emissions and deep sea sediments containing
Be$^{10}$, and suggested that it could be an extension of the Loop I
or North Polar Spur seen in the {Scorpius-Ophiucus} region. In the
opposite direction one study predicts that the supernova forming the
Antlia remnant exploded 1.20 Myr ago and 128 pc away at
$\ell=270.4,~b=19.2^\circ$ \cite{Tetzlaffetal:2013}.  Smith and Cox
(2001, \cite{sc01}) have shown that multiple SNe
within about 3 Myr can produce a bubble with conditions that resemble
the Local Bubble. However a model where a homogeneous local plasma at
temperatures $\sim 10^6$ K accounts for all of the low energy X-ray
emission has been elusive, as discussed in detail by Welsh and Shelton
(2009, \cite{ws09}).

No signs of a cluster of massive stars interior to the Local Bubble
have been found. Using a kinematic analysis of the entire solar
neighborhood within about 400 pc, Fuchs et al.~ (2006,
\cite{fuchsetal06}) have suggested that the youngest associations in
the solar neighborhood entered the present bubble region about 10-15
Myr ago, and that approximately 14-20 have exploded since then, a view
consistent with the earlier studies of Maiz-Apellaniz (2001,
\cite{Maiz-Apellaniz_2001}).  With the help
of non-equilibrium ionization modeling, de Avillez and Breitschwerdt
\cite{dab12} have constrained the evolution time since the last SN to
be about 0.5-0.8 Myr. These parameters are in rough agreement with
those derived from \fesixty\ (see Section  \ref{sec:fe60}).

Charge exchange (CEX) between the solar wind ions and interstellar
atoms has been suggested as at least a partial, if not complete,
source for the diffuse X-ray background by Cravens et al. (2001,
\cite{Cravens_etal_2001}).  The origin of foreground contamination near
0.75 keV and 0.25 keV differ, with the former primarily due to CEX
with solar wind oxygen atoms and the latter due to CEX with L-shell
states for many species for which transition strengths are unknown
\cite{KuntzSnowden:2015}.  Predictions of foreground CEX emission in the 0.75 keV
band yield a hard spectrum and does not predict the CEX rates
required to discount an interstellar source of the SXRB
\cite{Frisch:2009ibex,ws09}.  At lower energies, X-ray shadows are
seen in the 0.094 keV Wisconsin band.
The best models of simultaneous solar wind foreground
and a thermal {Local Bubble} hot plasma indicate that the Local Bubble
produces produces $26\% \pm 4\%$ of the 0.1--0.4 keV emissions
\cite{SmithFosterBrickhouse:2014}.  
An alternative analysis finds solar wind
charge exchange foreground levels of 43\% to 76\% of the SXRB
produced by the Local Bubble in the 0.25 keV ROSAT band in the
direction of the local cold Leo cloud \cite{snowden15a, snowden15b}.
These results seem to reaffirm that the Local Bubble cavity is filled
with a uniform hot gas, but also indicate that more efforts to
understand the foreground are needed.

The properties of the Local Bubble, and the similarity to other
(super)bubbles, coupled with the general observations of a 3-phase
interstellar medium with a hot phase consisting of low density, high
temperature gas \cite{cox05}, hint at a massive star and/or supernova
origin. Other evidence within the Local Bubble also points towards a
SN origin. \cite{nehmeetal08a} and \cite{gryetal09} study the peculiar
characteristics of a cometary-shaped infrared cirrus cloud towards the
star HD 102065. The interpretation attributes the spatial structure in the cold
phase, the high (and negative) velocities, the high abundance of atoms
in excited states, a high level of ionization associated with the
highest velocities, as well as the unusually high abundance of small
dust particles, as all due to the interaction of the molecular cloud
with a SN shock wave approximately 200,000 to 300,000 years ago.

\subsection{Loop I and the Very Local Interstellar Medium } \label{sec:loopI}

\begin{figure}[h!]
\begin{center}
\includegraphics[scale=0.65]{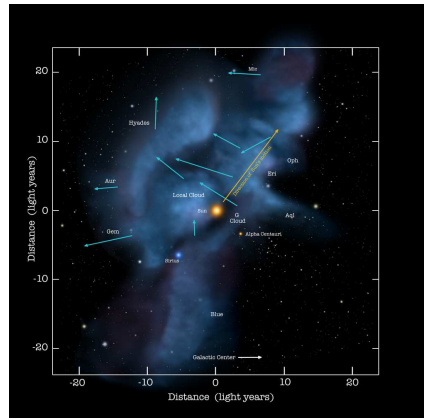}
\end{center}
\caption{{Nearby clouds:} Locations of the tenuous local interstellar clouds within $\sim 15$ pc
are shown projected onto the galactic plane, together with the cloud motions through the LSR 
(blue arrows, \cite{FrischSchwadron:2014icns}).  Names are shown for the clouds \cite{RLIV:2008}
and several nearby stars.  The solar apex motion through the LSR is marked with the yellow arrow. The center of the S1 shell is
about 78 pc beyond $\alpha$ Cen. Figure credit:  NASA, Adler Planetarium, P. C. Frisch, S. Redfield.}
\label{fig:clic}
\end{figure}

The physical characteristics of the nearest interstellar material,
including the geometry of Loop I, local cloud kinematics, interstellar
magnetic field, and gas-phase abundances, indicate an origin for the
cluster of local interstellar clouds (CLIC) within $\sim 15$ pc that
is related to the Loop I superbubble. An alternate origin for the CLIC
as a magnetic flux tube that detached from the Local Bubble walls (Cox
and Helenius, 2003 \cite{CoxHelenius:2003}) has not yet been tested
against recent data. The CLIC contains kinematically defined
interstellar clouds that are located mainly within 15 pc of the Sun
(Figure \ref{fig:clic}).  The result that the Sun is in the rim of the
Loop I superbubble rests on several arguments.

{\it Geometrical considerations:} Studies of the Loop I geometry
consistently place the Sun in or close to the rim of the Loop I bubble
if it is spherical.  Models of radio Loop I and the NPS as a single
shell yield a shell center at $\ell =325^\circ,~ b=17.5^\circ $, and
$130 \pm 75$ pc with a diameter of $230 \pm 135$ pc
\cite{Berkhuijsen:1971,Berkhuijsen_etal_1971,Spoelstra:1972lb,
  Berkhuijsen:1973r-I}.  Wolleben (2007,\cite{Wolleben:2007}) fit two
separate spherical superbubble shells, S1 and S2, to the polarized
Loop I radio continuum {emission} using a model of a spherical
superbubble shell {expanding} in a uniform magnetic field
(\cite{Frisch:2010s1}, Fig. \ref{fig:lb}, Fig. \ref{fig:santos},
left).  The Sun is located in the rim of the S1 shell that is centered
at $\ell \sim 346^\circ$ and distance $78 \pm 10$ pc, with the
distance comparable to the shell radius (62--101 pc).  The \HI\ 21-cm
shell of Loop I is centered at at $\ell = 320^\circ,~b=5^\circ$, and
has a distance and radius of $\sim 118$ pc \cite{Heiles:1998whence}.

{ \it CLIC kinematics:} The upwind direction of the bulk motion of the
CLIC through the LSR is directed toward the center of Loop I,
indicating that the Sun is in the Loop I shell that is still expanding
at a velocity of $ 17.3 $ \kms\ \cite{Frisch:2015ismf3}.  The
kinematics of the low density interstellar gas within 15 pc have been
evaluated using two different assumptions; as a coherent flow of
interstellar gas and dust through space
\cite{Crutcher:1982,FrischYork:1986,Bzowski:1988,FGW:2002,GryJenkins:2014clic},
or as a group of separate cloudlets with different velocities
\cite{Herbig:1968,Lallement:1986,LallementBertin:1992,FGW:2002,Frisch:2003apex,RLIV:2008}.
Cloud velocities are found from interstellar absorption lines
(e.g. \cite{Herbig:1968,RogersonYork:1973ions}). {Velocities} for the
cloud around the heliosphere {can also be estimated} from in situ
measurements of interstellar gas
(e.g. \cite{Moebius:2004he,Schwadron:2015He}) and dust
\cite{Frisch:1999,KimuraMann:2003velorigin} that share similar
velocities.  The upwind direction of the CLIC interstellar wind
\cite{Frisch:2015ismf3} coincides with the center of the S1 shell
\cite{Wolleben:2007}, with an angle of $14^\circ \pm 18^\circ$ between
CLIC LSR velocity and the S1 shell center.  The flow of local
interstellar gas away from the Loop I region was discovered decades
ago
\cite{Frisch:1979,frisch81,Crutcher:1982,FrischYork:1986,Bzowski:1988}
and recent fits to different selections of interstellar absorption
line data lead to similar results for the bulk flow; note that the
vector directions in Gry and Jenkins (2014,
\cite{GryJenkins:2014clic}) and Frisch et al. (2002, \cite{FGW:2002})
differ by $11^\circ \pm 17^\circ $.

{\it Perpendicular relation between interstellar magnetic field
  direction and LSR cloud velocities:} Swept-up field lines near the
equator of an expanding superbubble shell are perpendicular to the
expansion velocity \cite{FerriereMacLow:1991,HanayamaTomiska:2006}.
Wolleben \cite{Wolleben:2007} assumed such a configuration when
evaluating the geometry of the S1 and S2 shells.  This property gives
cloud velocities that are perpendicular to the ISMF for the equatorial
regions of the bubble.  The bulk velocity of the CLIC through the LSR
is perpendicular to the CLIC ISMF direction, and the LSR vector of the
Local Interstellar Cloud (LIC) gas that surrounds the heliosphere is
perpendicular to the {LIC} magnetic field direction.  The angle
between the bulk CLIC velocity and the interstellar magnetic field
direction determined from polarized light from nearby stars is
$80^\circ \pm 8^\circ$ \cite{Frisch:2011araa,Frisch:2015ismf3}.  The
ISMF direction is obtained from the statistical analysis of {polarized
  starlight}, giving an ISMF pole toward
$\ell=36.2^\circ,~b=49.0^\circ ~ (\pm 16^\circ )$
\cite{Frisch:2015ismf3}, where the polarization is caused by the
attenuation of starlight by a dichroic screen of foreground dust
grains aligned with respect to the ISMF \cite{Andersson:2015araa}.
The LIC magnetic field direction is found from the center of the IBEX
ribbon arc of higher ENA fluxes, $\ell=34.8^\circ \pm 4.3^\circ$,
$b=56.6^\circ \pm 1.2^\circ$, which forms upwind of the heliopause
where the ISMF draping over the heliosphere becomes perpendicular to
the sightline
\cite{McComas:2009sci,Schwadron:2009sci,Zirnstein:2016ismf}.  The
velocity of the LIC gas has been determined from IBEX in situ
measurements of neutral interstellar He, H, and O flowing through the
heliosphere
\cite{Moebius:2009sci,Schwadron:2014sci,Schwadron:2015H,McComas:2015isn}
and corresponds to a LIC LSR velocity of $17.2 \pm 1.9$ \kms\ toward
$\ell = 141.1^\circ \pm 5.9^\circ,~b=2.4^\circ \pm 4.2^\circ$
\cite{Frisch:2015ismf3}.  The LIC ISMF direction and LSR velocity are
nearly perpendicular, with an enclosed angle of $87.6^\circ \pm
3.0^\circ$ \cite{Schwadron:2014sci}.  The ISMF directions in the CLIC
and LIC agree to within $7.6 ^{+14.9} _{-7.6} $ degrees
\cite{Frisch:2015ismf3}.

{\it Local ISMF orders kinematics of local clouds:} An alternate view
of CLIC kinematics is provided by parsing the observed velocity
components into individual cloudlets. A self-consistent analysis of
separate cloud velocities has been developed by Redfield and Linsky
(2008, \cite{RLIV:2008}).  Comparisons between the LSR velocities of
these clouds \cite{FrischSchwadron:2014icns} and the IBEX ISMF
direction \cite{Funsten:2013,Frisch:2015ismf3} reveal that the LSR
cloud velocities are roughly proportional to the angle between the LSR
cloud velocity vector and the ISMF direction (see Figure 12 of Frisch
et al.~(2015, \cite{Frisch:2015ismf3}).

{\it Abundance pattern of gas-phase elements:} Clear evidence that the
local ISM has been processed {by passage} through supernova
shocks is provided by {comparisons between} the abundance patterns
of interstellar gas and solar abundances. Elements
missing from the gas are due to the condensation of minerals onto dust grains
\cite{Ebel:2000}.
Frisch \cite{frisch81} and Crutcher \cite{Crutcher:1982} attributed the 
relatively high abundances of refractory elements in nearby
interstellar gas to the erosion of
grains by shocks originating in the {Sco-Cen
Association}.  Early data on Ca II and Na I gas-phase {abundances} found
that the abundances of Ca II and other refractory elements
increased with the cloud {LSR} velocity
\cite{RoutlySpitzer:1952,Spitzer:1976,SembachDanks:1994} because of
the processing of interstellar dust in high-temperature shocks that
erode the refractory component of the grains
(e.g. \cite{SilukSilk:1974,Jones:1996,Frisch:2009ibex,Slavin_etal_2015}.
Interstellar depletions for 243 sightlines (Jenkins 2009,
\cite{Jenkins:2009}) have been characterized by considering the common
parameter that describes the {depletion} pattern as a function of
element, and a second parameter that describes depletion patterns between
sightlines.  Interstellar depletions increase with the total hydrogen column
density, but unrecognized ionization of the gas will produce inaccurate
weaker depletions of refractory elements such as
\FeII\ and \MgII\ that have low first ionization potentials.
The abundance patterns in the CLIC gas are
similar to those of warm clouds \cite{Welty:199923ori} but vary
between individual cloudlets \cite{RLIV:2008}.  The velocities of LIC
gas \cite{Schwadron:2015He} and LIC {dust} (from in situ measurements of
interstellar dust inside of the heliosphere
\cite{Frisch:1999,KimuraMann:2003velorigin}) are in agreement
indicating that the grain destruction must have occurred far in the
past.

{\it The best constrained interstellar cloud, the LIC, has an origin
  in a superbubble shell:} The best understood interstellar cloud is
the LIC that feeds interstellar gas
\cite{Frisch:2009ibex,Frisch:2011araa,McComas:2015six} and dust
\cite{Frisch:1999,KimuraMann:2003velorigin,Sterken:2015sixteen,Altobelli:2016cassini}
into the heliosphere, and is detected toward over 75 stars
\cite{RLIV:2008}.  IBEX data permit the detailed study of the LIC at
one spatial location; those data show that the LIC LSR velocity and
LIC ISMF directions are perpendicular and the upwind direction of the
LIC velocity is toward the center of Loop I
\cite{Schwadron:2014sci,McComas:2015isn,Frisch:2015ismf3}.  Abundances
in the LIC have been corrected for ionization effects using
self-consistent radiative transfer models \cite{SlavinFrisch:2008}.
Components of the models include a source of EUV photons to account
for high ratios of \HI/\HeI\ found in pickup ion and anomalous cosmic
ray data inside of the heliosphere
\cite{Fisk+Ramaty_1974,Gloeckler_Fisk_2007,Witte:2004,CummingsStone:2002},
and toward nearby white dwarf stars
\cite{Frisch:1995rev,Wolffetal:1999}.  Pickup ions and anomalous
cosmic rays form from interstellar neutrals that survive penetration
into the heliosphere \cite{Bzowski:2013neutralsurvival} and are either
directly sampled through in situ measurements
\cite{McComas:2009sci,Moebius:2009sci,Park:2014NeO,Schwadron:2015H,McComas:2015isn,Park:2015isn,Schwadron:2016O}
or ionized through {charge-exchange}, {photoionization}, and other
processes \cite{Bzowski:2013neutralsurvival} to create the pickup ion
population \cite{Fisk+Ramaty_1974,Gloeckler_Fisk_2007} or accelerated
to become the anomalous cosmic ray population
\cite{CummingsStone:2002}.  Using LIC data toward $\epsilon$ CMa and
pickup ion and in situ heliospheric data, elemental abundances have
been reconstructed for the LIC
\cite{Slavin:1989,SlavinFrisch:2002,SlavinFrisch:2008}. {Predictions
  from these models include} densities of \nHI$\sim 0.19$ \cc,
\nel$\sim 0.07$ \cc, ionization levels of hydrogen and helium $\sim
22$\% and $\sim 39$\% {respectively} for a cloud temperature of 6300
K, the full destruction of {carbonaceous grains} in the LIC, and
elevated {gas-phase abundances} for Fe and Mg that indicate silicate
grains \cite{SlavinFrisch:2008}.  In situ measurements of interstellar
HeI inside of the heliosphere yield a LIC temperature of $8000 \pm
1300$ K \cite{Schwadron:2015He}.  Solar abundances are also found for
carbon in the low density gas at intermediate and high velocities
toward Orion \cite{Welty:2002zeta,Welty:199923ori}.  The LIC abundance
pattern fits into the interstellar abundance patterns that depend on
cloud velocity, which are nicely established for the low-velocity,
intermediate velocity, and {high-velocity clouds} studied towards
towards Orion where different clouds have been {shocked} differently
(Welty et al.~ 1999, 2002, \cite{Welty:199923ori,Welty:2002zeta}).

\subsection{Line-of-sight Blending of Loop I and Loop IV with Galactic Center Backgrounds}
Several recent studies \cite{Sofue:2015nps,Lallement:2016nps} have
attributed X-ray features toward Loop I to the gamma-ray bubbles
around the galactic center found by Fermi-Lat
\cite{SuFinkbeiner:2010}.  This hypothesis requires that the
Fermi-bubbles extend over 5 kpc into the galactic halo. The arguments
for a galactic center origin of Loops I and IV rely partly on the
latitude dependence of the North Polar Spur X-ray emission compared to
that of dust, \HI\ or molecular material.  The North Polar Spur was
originally defined as an intense source of non-thermal radio
continuum, and subsequently associated with Loop I and found be a
strong source of X-ray emission.  While distant contributions to the
North Polar Spur soft X-ray emission can not be ruled out, especially
given the complex spectrum for the NPS X-ray emission at $\sim 0.15$
keV suggestive of a reheated supernova remnant
\cite{Iwan:1980npsXrayloopI}, studies of polarized stars with known
distances clearly prove that the main radio continuum Loop I is a
local phenomena within $\sim 200$ pc
\cite{Santos:2011,BerdyuginPiirola:2014s1}.  Faraday tomography of the
radio continuum adjacent to the North Polar Spur region indicates that
radio emission from the spur is not Faraday-depolarized and therefore
most likely within a few hundred parsecs \cite{Sun:2015faratomnps}.
Foreground and background structures are difficult to distinguish when
the shadowing interstellar material consists of magnetically organized
dust structures with sub-parsec filaments and collapsing clouds. The
Aquila Rift set of molecular clouds shadows the North Polar Spur X-ray
emission \cite{Sofue:2015nps,Lallement:2016nps} but do not negate the
optical polarization data that show a local origin for the Loop I
magnetic field \cite{Santos:2011,BerdyuginPiirola:2014s1}.

Loop I and Loop IV are prominent high-latitude radio bubbles in
galactic quadrants IV and I.  Berkhuijsen et
al. (1971, \cite{Berkhuijsen_etal_1971}) interpret these loops as
supershells associated with evolved supernova remnants.  The centers
and diameters for Loop I and Loop IV are, respectively
$ \ell, b =329^\circ \pm 1.5^\circ ,+17.5^\circ  \pm 3^\circ$, $116 \pm 4$ pc,
and $\ell, b =315^\circ \pm 3^\circ , +48.5^\circ \pm 1^\circ$,
$39.5 \pm 2 $ pc.  The galactic bulge is 8.5 kpc beyond Loop I.
Reynolds (1984, \cite{Reynolds:1984alpvir}) has shown that Loop IV is
associated with a large hole in the distribution of nearby
interstellar neutral hydrogen that coincides with an extended region
of ionized hydrogen visible through H$\alpha$ emission, and
surrounding the hot variable star Spica ($\alpha$ Virginis,
$80.8 \pm 6.9$ pc, B1V).  The quasar 3C273 is viewed through the rim
regions of Loop I and of Loop IV leading to complications in the
interpretion of highly ionized gas in the X-ray spectrum of
3C273 \cite{FangJiang:2014_3C273}.  Ultraviolet observations
of the halo star HD 119608, located at 4.1 kpc and beyond Loops I and IV, show the
bimodal velocity structure of an expanding shell \cite{Sembach:1997LoopI_IV}. The distance of HD 119608 
and its foreground expanding shell indicate that the large-scale Loop I does not
originate in the galactic center, while the coincidence of the HII region around Spica and Loop IV 
indicate that Loop IV is local.

\subsection{The Orion-Eridanus Superbubble} \label{sec:ori}
\begin{figure}[h!]
\begin{center}
\hspace*{-0.1in}
\includegraphics{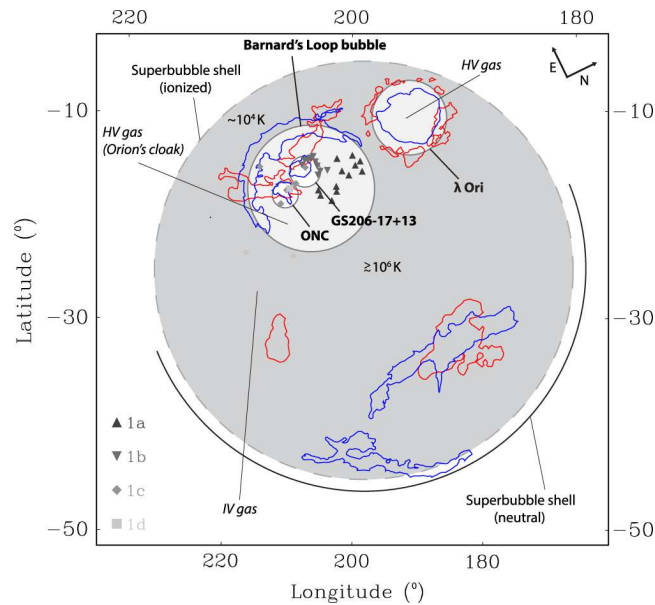}
\end{center}
\caption{ {Schematic of the {Orion-Eridanus} superbubble superbubble and
    several of its major components \cite{obbt15}:} The Orion-Eridanus
  superbubble is shown together with the structures that trace this
  bubble.  A superbubble formed from SNe in an old subpopulation of
  the Orion OB association and is traced by intermediate-velocity (IV)
  shocks \cite{Welty:199923ori,Welty:2002zeta}.  At lower galactic
  latitudes the superbubble is surrounded by a shell of neutral
  swept-up material (solid black line) that is not apparent in the
  opposite direction. Nested younger and smaller shells are shown
  (solid gray circles) such as the famous Barnard's Loop feature.  See
  Ochsendorf et al.~(2015, \cite{obbt15}) for more information.}
\label{fig:orion}
\end{figure}

The Orion region is the closest region 
that is actively forming high-mass stars. The activity of all the
stars has combined to create the Orion-Eridanus bubble. At a distance
of $\sim$ 400 pc, it is a nearby expanding structure, explored over
the entire wavelength range \cite{hhr99}, that serves as a testbed for
superbubble theories. A recent paper \cite{obbt15} has attempted to
synthesize the previous data along with data collected from WISE and
Planck to create a more complete model of the superbubble. The general
picture emerging from their investigations is that the bubble is
larger and more complex than was previously assumed (Figure \ref{fig:orion}). It consists of a
series of nested shells, the youngest of which, around the Orion
Nebula Cluster, is less than 1 Myr old. Some other smaller bubbles
triggered by ongoing activity are found around $\lambda$ Ori, and the
bubble GS206-17+13, most likely a stellar wind bubble approximately
centered on the $\sigma$ Ori cluster. In this model, Barnard's Loop is
part of a complete bubble structure associated with a SNR that
exploded about 0.3 Myr ago and then connected with the high-velocity gas
detected in absorption studies towards this region. The expansion
velocity of this structure is quite high, on the order of 100 km
s$^{-1}$, again suggesting a recent origin. The outer shell of the
Orion superbubble can be traced by observations of the intermediate
velocity gas towards this region. In the west, some remnant of the
neutral dense shell of the bubble can be seen, whereas towards the
east the shell is completely ionized. High temperature ($\ge 10^6$ K)
X-ray emitting gas is seen towards the west, while the gas temperature
in the eastern interior is two orders of magnitude lower.

The presence of various nested bubble structures suggests periods of
episodic star formation in the Orion-Eridanus region. Instead of
continuous input forming a single superbubble, each burst of star
formation gave rise to different subgroups of stars that may locally
ionize their environment, while stellar winds and the resulting SN
explosions modify and tend to sweep up the surrounding medium into a
bubble-like structure. The superbubble itself is probably 5-10 Myr old
and was formed by a series of {SNe} that arose from stars in the Orion
OB association. This model is supported further by the work of (Pon et
al., 2016 \cite{ponetal16}), who have attempted to fit a Kompaneets
model of a superbubble expanding in an exponential atmosphere to this
picture of the Orion-Eridanus bubble. They find morphologically
consistent models with reasonable Galactic scale heights of 80 pc,
provided that the bubble is oriented with the Eridanus side (at lower
latitudes) further from the Sun than the Orion side.

\section{Isotopic and Abundance Indicators of SN Activity} \label{sec:iso}

\subsection{$^{60}$Fe radioisotope Observations as Indicators of Nearby Supernovae} \label{sec:fe60}
Measurements of $^{60}$Fe in the terrestrial geological record provide
an amazingly good indicator of encounters between supernovae ejecta
and the heliosphere during the past several million years.  \fesixty
is a radioactive isotope of iron with a half-life of 2.62 million
years \cite{rugeletal09}. It is primarily produced in core-collapse of
massive stars, which typically eject 10$^{-5}$ to 10$^{-4}$
M$_{\odot}$ of \fesixty \cite{lc03}.  A small amount may be produced
during the s-process before core-collapse, or by {Asymptotic Giant
  Branch} (AGB) stars. The important point is that there are no
natural, terrestrial methods that produce {\fesixty}; therefore any
terrestrial reservoirs of \fesixty must be generally attributed to
earlier deposition due to core-collapse supernovae (SNe), and can be
considered as a signpost of the imprint of a nearby SN.

Using accelerator mass spectrometry, \cite{knieetal99} found evidence
of enhanced concentrations of \fesixty radioactivity in deep ocean
ferromanganese crust in the South Pacific. Further and better
measurements led them to suggest the presence of a significant
increase in the \fesixty concentrations about 2.8 Myr ago
\cite{knieetal04}, suggesting the presence of a SN explosion within a
few tens of parsecs from the solar system. Fields et al.~\cite{fhe05}
combined the data with SNe nucleosynthesis models to refine the
distance of a probable nearby SN to between 15 and 120 pc. Basu et
al. (2007, \cite{basuetal07}) posited an alternate theory, that the
\fesixty was due to the presence of micrometeorites trapped by the
crust rather than injection by a SN, but many of their arguments were
refuted by \cite{fitoussietal08}.

Recent work seems to further substantiate the SN origin of
\fesixty. Wallner et al.~ (2016, \cite{wallneretal16}) found that the
\fesixty signal was global by finding evidence of \fesixty deposition
in deep-sea archives from the Indian, Pacific and Atlantic
oceans. Furthermore, they find interstellar influx \fesixty onto earth
via dust grains between 1.7-3.2 Myr ago, with a second signal 6.5-8.7
Myr ago. They argue that these signals suggest recent massive star
ejections, presumably {supernova explosions}, in the solar
neighborhood within about 100 pc. Breitschwerdt et al. (2016,
\cite{bfsetal16}) have modeled the SN explosions that created the
Local Bubble based on the evolution and supernova rates in star
clusters forming the Sco-Oph groups.  They suggest that the \fesixty
signal is mainly due to two SNe that occurred 1.5 and 2.3 Myr ago, and
between 90 and 100 pc distance from the solar system. The progenitor
stars were about 9 M$_{\odot}$.  These calculations assumed that the
stars in the clusters were co-eval, which as pointed out by
\cite{Pecaut2012} and \cite{Pecaut2016} may not be the case. This
could result in a modification of the mass of the progenitor star.  An
8-10 M$_{\odot}$ SN occurring 2.8 Myr ago, with material arriving at
the Earth 2.2 Myr ago, was estimated by Fry et al.~ (2016,
\cite{ffe16}).

Deep-ocean crusts are not the only evidence of \fesixty
concentrations. Fimiani et al.~ (2016, \cite{fimianietal16}) have
confirmed earlier measurements \cite{cooketal09,fimianietal12} that
showed an excess of \fesixty concentrations in lunar cores, which
presumably originated from the same events that led to the \fesixty
deposition in ocean crusts. By measuring the concentration of
$^{53}$Mn in the same samples, they suggest that the \fesixty is
likely of SN origin, and that SN debris arrived on the moon about 2
Myr ago.

The Earth's {microfossil} record includes \fesixty of biological origin.
In 2013, Bishop et al.~\cite{blec13} analyzed Pacific ocean sediment drill
cores. They were able to extract \fesixty from magnetofossils
and quantify abundances using a mass spectrometer. Further analysis recently reported by
\cite{Ludwigetal16} confirms the direct detection of live \fesixty
atoms contained within secondary iron oxides, including
magnetofossils, which are fossilized chains of magnetite crystals
produced by magnetoactive bacteria. They suggest that the \fesixty
signal begins 2.6 - 2.8 Myr ago, peaks around 2.2 Myr earlier and
terminates around 1.7 Myr earlier, consistent with the time periods
deduced from other data such as deep-ocean crusts and lunar samples.

The composition of galactic cosmic rays reveals their origin as well
as provides hints to the {cosmic ray acceleration} mechanisms.  Using
the Cosmic Ray Isotope Spectrometer (CRIS) instrument on the
Advanced Composition Explorer (ACE) spacecraft, \fesixty has been detected in cosmic rays
of a few hundred MeV/nucleon \cite{binnsetal16}. The
\fesixty/$^{56}$Fe source ratio is (7.5 $\pm$ 2.9) $\times 10^{−5}$,
which is consistent with that produced in massive stars. The detection
of SN-produced \fesixty in cosmic rays indicates that the time
required for the acceleration and transport of the cosmic rays to
earth cannot exceed the half-life of \fesixty of 2.62
Myr. Consequently the distance from the source should be comparable to
the distance that cosmic rays can diffuse over this time period, which
they estimate to be less than 1 kpc. This is {consistent with}
the existence of a SN within a kpc {that exploded} during the
last 2.6 Myr.

\subsection{Isotopes and the OB Association Origin of Galactic Cosmic Rays} \label{sec:gcr} 

Elemental abundances of galactic cosmic rays (GCRs) reveal both the origin
and the acceleration mechanism of the GCRs.  The largest differences
between the isotopic composition of the GCRs and solar system values
are found for the ratios \neontwentytwo/\neontwenty,
\carbtwelve/\oxysixteen and \fefeight/\fefsix.  The ratio
\neontwentytwo/\neontwenty in GCRs is a factor of $5.3 \pm 0.3$ larger
than the value in the solar wind \cite{Binns:2005}.  Measurements of
\neontwentytwo/\neontwenty in the anomalous cosmic ray population
(ACR) at lower energies, which form from interstellar neutrals that become
charged while interacting with
heliospheric plasma, show ratios consistent with solar values
\cite{Leske:2000acrisotope}.  The enhanced ratio
{\neontwentytwo/\neontwenty} in GCRs suggests a source that includes
contributions from the ejecta of massive stars. Binns et al. (2005, \cite{Binns:2005}) 
have shown that the ratio can be explained as resulting from a mixture of
$\approx$20\% massive star ejecta and wind material with $\sim$80\%
interstellar medium. In fact they further show that such an admixture
could explain a range of isotope and element ratios for Z $\le$ 28
nuclei. Newer measurements \cite{MurphyBinns:2016} of abundances with
Z $\ge$ 26 are consistent with this, and indicate that GCRs have
formed from a mixture of $19$\% ejecta from massive stars and $81$\%
interstellar material with solar system composition. This means that
the stellar winds and/or supernova have mixed with only about 4 times
their ejected mass, which implies that the stellar source must not be
mixing with too much interstellar material, suggesting the nearby
presence of massive stars and/or supernovae.

Another isotopic constraint from GCRs is the lack of \nickelfn in
cosmic rays. {\nickelfn} has a half-life of 76,000 years before it
decays to {\cobaltfn} by electron capture. Once it is accelerated to high
energy, the \nickelfn is stripped off electrons, and therefore cannot
decay. Data from the CRIS experiment show a lack of \nickelfn in GCRs,
indicating that the \nickelfn has decayed from the amount one would
expect from a SN explosion \cite{wbc99}. Therefore Wiedenbeck et al.~ (1999, \cite{wbc99})
suggest that acceleration of the material took place at least 76,000
years after it was ejected. Thus while the \neontwentytwo/\neontwenty
ratio suggests that acceleration could not have taken place more than
a few million years after the wind material was ejected (to avoid
mixing too much interstellar medium material), the \nickelfn
measurement shows that it must be at least 0.1 Myr after the SN
explosion. It is possible that acceleration in superbubbles can
satisfy both these constraints. However, it should be noted that a
recent analysis by Neronov and Meynet (2016, \cite{nm16}), taking the yield of \nickelfn from
updated massive star models for stars up to 120 $\msun$, suggests a low
\nickelfn yield compared to \cobaltfn, consistent with the CRIS
  experiments, which may remove the need for this constraint.

Balloon-born {TIGER} measurements of heavy GCRs \fetsix through
\ziforty \cite{Rauch:2009gcrdust,MurphyBinns:2016} show that the
abundances of \fetsix through \ziforty in the galactic cosmic ray
population adhere to a pattern where the refractory elements in the
GCR population are enhanced over volatile elements. Refractory
elements in GCRs with energies of hundreds of MeV per nucleon to GeV
per nucleon show a preferential acceleration of a factor of four over
the acceleration of volatiles \cite{MurphyBinns:2016}. Epstein (1980,
\cite{epstein80}) was one of the first to explain the high abundance
of refractory elements in cosmic rays, by suggesting that they were
preferentially accelerated. Bibring and Cesarsky (1981, \cite{bc81})
indicated that SN shocks can pick up particles from a suprathermal
population produced by the destruction of dust grains.  Ellison et
al.~ (1997, \cite{EllisonDruryMeyer:1997}) produced the first detailed
model explaining GCR abundances and isotopic ratios using interstellar
grains that are accelerated to modest energies by SN shock waves.  Grain
destruction in the shock layer results from the thermal sputtering of
particles from grain surfaces due to gas-grain collisions and grain
shattering and vaporization during grain-grain collisions
\cite{BarlowSilk:1977,DraineSalpeter:1979a,Jones:1994,Jones:1996}.
{Refractory elements show a well-known resilience against
  destruction. High condensation temperatures lead to refractory
  elements being injected into the shock at higher energies than
  volatiles, giving rise to the preferential acceleration of
  refractory elements in contrast to the volatiles that are
  accelerated in accordance with their mass-to-charge ratio (Ellison
  et al. 1997, \cite{EllisonDruryMeyer:1997}).}

Signatures of a nearby supernova have been postulated in the locally
observed cosmic ray spectrum. \cite{Kachelriessetal15} have suggested
that the excess of positrons and antiprotons above $\sim$20 GeV, and
the discrepancy of slopes in the spectra of cosmic ray protons and
heavier nuclei in the TeV-PeV energy range, can be explained as due to
a nearby source, which was active about 2 Myr ago. This source
injected about 2-3 $\times 10^{50}$ erg of energy in cosmic rays. The
transient nature and overall energy budget suggest a SN origin, with
an age equal to that given by the other indicators above.

\subsection{\altwentysix\ as a Tracer of Massive Stars}

In 1999, Knodlseder~\cite{knodlseder99} showed from an analysis of
Comptel data that the 1.8 MeV gamma-ray line was closely correlated
with the 53 GHz free-free emission in the Galaxy. 1.8 MeV gamma-rays
are emitted during the radioactive decay of \altwentysix, which has a
half-life of about 0.7 Myr. The free-free emission arises from the
ionized interstellar medium. He argued that this could be understood
if massive stars are the source of {\altwentysix}. Knodlseder et
al.~\cite{knodlsederetal99} showed that the correlation was also
strong with other tracers of the young stellar population, which
confirmed that the source of \altwentysix\ was massive stars and
supernovae.

Using spatial maps from the {Comptel} observatory to identify isolated
regions of $\gamma$-ray emission, and the {INTEGRAL} $\gamma$-ray
spectrometer to identify the $\gamma$-ray velocities,
Diehl et al.~(2010, \cite{diehletal10}) identified a $\gamma$-ray source expanding toward
the Sun at $137 \pm 75$ \kms\ from a 10$^\circ$-radius region centered
on the {Upper-Scorpius} (US) subgroup of the Sco-Cen Association.  Given
the 0.717 Myr half-life of \altwentysix, this implies that the massive stars
were born less than 10 Myr ago, thus indicating recent star formation.
De Geus (1992, \cite{deGeus:1992}) suggested that the proto-US cloud
was compressed by an expanding shell from the Upper Centaurus-Lupus
association $\sim 4$ Myr ago, igniting star formation.  Since the
velocity of the $\gamma$-ray source exceeds the $\sim 10$
\kms\ velocity of the \HI\ shell around the US subgroup, Diehl et
al. \cite{diehletal10} have adopted the scenario where the high-velocity gas is stellar
ejecta streaming into an adjacent preexisting cavity, and that
deceleration occurred as the gas collided with the preexisting walls
of the bubble \cite{diehletal10}.  The
young star clusters in a spiral arm will feed \altwentysix\ ejecta into
pre-existing \HI\ supershell cavities that were left over from 
earlier star formation during passage of the
previous spiral arm density wave \cite{KrauseDiehl:2015}.

\section{Impact of Supernova on Heliosphere} \label{sec:helio}

Supernovae impact the heliosphere through the direct encounter of the
heliosphere with the SN blast wave or ejecta, or modification of the
ISM properties at the heliosphere.  As discussed above, the Sun is
traveling through the shell of the Loop I superbubble that resulted
from stellar evolution in the ScoCen OB2 association.  It was
recognized long ago that extreme variations in the physical properties
of interstellar material interacting with the solar system would
probably affect the terrestrial climate
\cite{Shapley:1921,HoyleLittleton:1939climate,McCrea:1977snrclimate,Frisch:1993gstar,Scherer:2006helioclimate,Frisch:2006book,MuellerFrisch:2006apj}.
These effects are mediated by the interaction between the heliosphere
and interstellar medium
\cite{Holzer:1989,Frisch:1993gstar,MuellerFrisch:2006apj,Scherer:2006helioclimate,ZankFrisch:1999,Zank:2015araa}.
The 18 \kms\ motion of the solar system through the LSR and the 7--47
\kms\ LSR velocities of nearby interstellar clouds
\cite{FrischSchwadron:2014icns} lead to variations in the
{heliospheric} boundary conditions over geological timescales of
order $\le$ 30 kyr \cite{FrischMueller:2011ssr}.  Implications of our
changing galactic {environment} are discussed in Scherer et al. (2006,
\cite{Scherer:2006helioclimate}) and Frisch (2006,
\cite{Frisch:2006book}).

The heliosphere configuration is governed by the relative pressures of
the solar wind and interstellar material, including the dynamic {ram
pressure} that increases non-linearly {with the interaction
  velocity} $\sim V^2$, and {by} interstellar ionizations since
excluded ions and penetrating neutrals interact differently
\cite{Holzer:1989,Zank:2015araa}.  Even a moderate increase in the
relative velocities of the Sun and surrounding interstellar cloud from
the present 25.4 \kms\ LIC velocity
\cite{Schwadron:2015He,McComas:2015isn} to 45 \kms\ (such as found for
the cloud named ``Vela'' by Redfield and Linsky 2008,
\cite{RLIV:2008}) leads to a decrease of the {heliopause} distance by
34\% percent from 104 AU to 69 AU \cite{MuellerFrisch:2006apj}.
Interactions between the heliosphere and an evolving superbubble at
different velocities leads to different configurations for the
heliosphere. Variations in the heliosphere-interstellar interactions
also arise from variations in the {solar magnetic activity cycle} (e.g.
\cite{WashimiTanaka:1999,PogorelovSuess:2013solarcycle}).

\begin{figure}[h!]
\begin{minipage}{14pc}
\includegraphics[width=14pc]{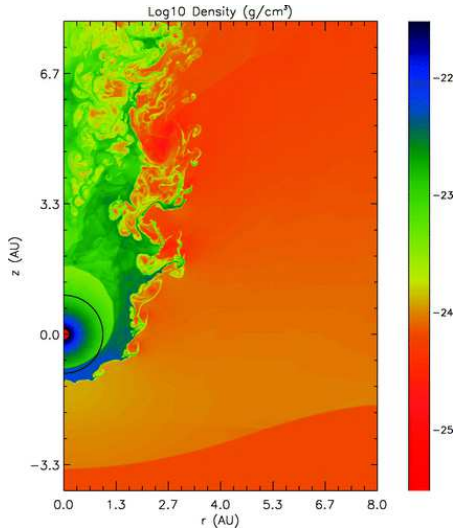}
\caption{\label{fig:sn} One scenario showing the impact of a SNe blast
  wave on the heliosphere.  A map of the logarithmic density of the
  heliosphere is shown after interaction with the shock from a
  supernova explosion 10 pc away for an explosion occurring in a
  tenuous unmagnetized low density cloud similar to the LIC.  The Sun
  is located at the origin.  Two shocks result, the innermost
shock forms where the solar wind decelerates from a supersonic
to a subsonic plasma, and the outer shock, the
  heliospheric bow shock, occurs where the 
  blast wave becomes subsonic. The interface between the two
  fluids is marked by Kelvin-Helmholtz instabilities.  In this model
  the SNR dominates the solar wind at 1 AU.  See Fields et al. (2008,
  \cite{Fields:2008snhelio}) for additional information. }
\end{minipage}\hspace{2pc}%
\begin{minipage}{14pc}
\includegraphics[width=14pc]{fig6b.eps}
\caption{\label{fig:gcr} Galactic cosmic ray fluxes at Earth for
  different solar environments.  Higher GCR fluxes are found at
  the Earth when the Sun is immersed in a fully ionized hot plasma
  (blue lines) compared to for today's immersion of the
Sun in the LIC
  (green lines).  This effect arises from the charge-exchange between
  the interstellar neutrals and the solar wind that increases
  turbulence and the modulation of the GCR component through
  mass-loading.  See Mueller et al. (2008,
  \cite{Muelleretal:2008time}) for more information.}
\end{minipage} 
\end{figure}
\vspace*{-0.11in}

The extreme example of the influence of our galaxy on the heliosphere
would arise from the explosion of a supernova close to the Sun.  The
wind from the massive-star preceding a {core-collapse supernova} would,
if close to the Sun, create a fully ionized environment for the
heliosphere.  The supernova explosion would first be noticed by an
intense flux of UV/X-ray photons from the SN explosion itself that
could ionize and heat the cloud around the heliosphere.  A
counter-intuitive result is that the flux of galactic cosmic rays at
the Earth will increase for immersion of the heliosphere in a fully
ionized plasma (Figure \ref{fig:gcr}).  Penetrating {interstellar
neutrals} become ionized through charge-exchange, photoionization,
and other processes
\cite{Bzowski:2013neutralsurvival} to create {pickup ions} that become
accelerated to form {anomalous cosmic rays} \cite{Fisk+Ramaty_1974}.
The pickup ions that are trapped on the solar wind magnetic field
lines mass-load the wind and increase turbulence that impedes the
propagation of galactic cosmic rays to the inner heliosphere.  Figure
\ref{fig:gcr} shows that GCR fluxes in the inner heliosphere could
increase by an order of magnitude if the surrounding interstellar
cloud became fully ionized by a nearby supernova.

A {blast wave} from a nearby supernova would compress the heliosphere.
Heliosphere multifluid models predict an encounter with a decelerated
super bubble shell, at a velocity 100 \kms\ relative to the
heliosphere, would shrink the heliopause to $\sim 14$ AU for a warm
tenuous cloud (8,000 K, $n \sim 0.8$ \cc)
\cite{MuellerFrisch:2006apj}.  Numerical simulations show that a blast
wave of thousands of \kms\ would sweep away the heliosphere, possibly
leaving the Earth directly immersed in the supernova remnant.  Fields
et al. (2008, \cite{Fields:2008snhelio}) simulated a scenario for the
heliosphere responding to a supernova that is located 8 pc away
(Fig. \ref{fig:sn}).  The interface between the solar wind and remnant
plasma becomes highly unstable due to Kelvin-Helmholtz
instabilities. These particular simulations do not explore a medium
{modified by} the winds of the progenitor star
(e.g. \cite{gml96a,vvd05,vvd07}), which could alter the scenario.

The proximity of active OB stars to the Sun {could lead to an
  excess of photons at the time of shock breakout} that {would
  result in} a potentially significant influence on the physical
conditions of the interstellar medium that shapes the heliosphere.
There is no a priori basis for assuming a constant radiation field at
the Earth over the past $\sim 1$ Myr.  Slavin and Frisch (1996,
\cite{SlavinFrisch:1996vela}) have modeled the {photon burst} created by
the supernova parent of the young {Vela pulsar}, $<30,000$ years old and
located at edge of the low density Local Bubble region in the near
side of the Gum nebula, and suggested that the supernova may have
contributed to the ionization of the interstellar cloud around the
heliosphere.  {Recombination} times for LIC-like gas at densities 0.1
\cc\ are $\sim 650,000$ years and longer if the gas is hotter.
Possible sources of the ionizing radiation include {$\gamma$-rays} and
the break-out of the supernova shock from the stellar atmosphere.
Brakenridge(1981, \cite{Brakenridge:1981}) concluded that radiation
from a nearby supernova may have left {isotopic} signatures in
the \cfourteen\ terrestrial record, and that the Vela supernova may
explain a \cfourteen\ anomaly 15,000 years ago.

The photobiological effects of a supernova that occurred 2.5 Myr ago,
at a distance of 50 pc, have been recently explored by
\cite{thomas17}. They conclude that biological impacts due to
increased UV irradiance by the nearby SN were not at a mass-extinction
level, but could have contributed to changes in the abundances of
various species. Such a conclusion is consistent with species turnover
observed around the Pliocene-Pleistocene boundary.

\section{Conclusions}

The interstellar medium of the solar neighborhood within $\sim 500$ pc
has been {shaped} by the massive stars assigned to Gould's Belt in
earlier studies.  The nearest region of star formation is in the
Scorpius-Centaurus Association where multiple supernovae have erupted
during the past 15 Myr.  Winds and supernova in these regions create
bubbles and superbubbles that remix the interstellar medium over
spatial scales of 500 pc.  These bubbles are detected {as} filaments
and loops of synchrotron emission arising from the compressed
interstellar magnetic fields in the bubble walls and/or shells of
\HI\ gas swept up by the expanding bubbles.  Two examples are the Loop
I superbubble, which has expanded to the solar location, and the more
distant Orion-Eridanus superbubble.  The cluster of local interstellar
clouds, as well as the ISM flowing through the heliosphere, display
{signatures} of an origin inside of the rim of the Loop I superbubble,
including the interstellar magnetic field direction and cloud
velocities through the LSR.  Cosmic ray isotopes trace the mixing of
local interstellar material with the interstellar medium.  {The
  discovery of short-lived radio-isotopes in the geologic record
  indicate that the Earth has received material from supernovae
  occurring within the past $\sim $2 Myr}.  Heliosphere models show
that the heliosphere is a sensitive barometer for interstellar
pressures and would react dramatically to the explosion of a nearby
supernova.  A diversity of astrophysical and geological data are
converging to allow new insights into the origin of interstellar
material around the Sun and to expand our perspective to include the
relation between the heliosphere and the Milky Way galaxy.

{\it Acknowledgments:} PCF would like to thank NASA for the support of
this research through the IBEX funding component of the NASA Explorer
program, and through a grant from the Space Telescope Science
Institute.  VVD is grateful for support from NASA Emerging Worlds
grant NNX15AH70G. His work on supernovae, circumstellar interaction
and wind-blown bubbles has been supported over the years by NASA and
the Chandra X-ray Center. VVD acknowledges useful discussions with the
members of Team 351 at the ISSI in Bern, Switzerland, April 2016.


\begin{thebibliography}{100}
\providecommand{\url}[1]{{#1}}
\providecommand{\urlprefix}{URL }
\expandafter\ifx\csname urlstyle\endcsname\relax
  \providecommand{\doi}[1]{DOI \discretionary{}{}{}#1}\else
  \providecommand{\doi}{DOI \discretionary{}{}{}\begingroup
  \urlstyle{rm}\Url}\fi

\bibitem{Shapley:1921}
H.~Shapley, J. Geology \textbf{29} (1921)

\bibitem{FlorinskiZank:2006jos}
V.~Florinski, G.P. Zank, in \emph{{Solar Journey: The Significance of Our
  Galactic Environment for the Heliosphere and Earth}} (Springer, 2006), pp.
  {281--316 }

\bibitem{MuellerFrisch:2006apj}
H.R. {M{\"u}ller}, P.C. {Frisch}, V.~{Florinski}, G.P. {Zank}, \apj,
  \textbf{647}, 1491 (2006)

\bibitem{Brakenridge:1981}
G.R. {Brakenridge}, Icarus, \textbf{46}, 81 (1981)

\bibitem{Frisch:2015ismf3}
P.C. {Frisch}, B.G. {Andersson}, A.~{Berdyugin}, V.~{Piirola}, H.O. {Funsten},
  A.M. {Magalhaes}, D.B. {Seriacopi}, D.J. {McComas}, N.A. {Schwadron}, J.D.
  {Slavin}, S.J. {Wiktorowicz}, \apj, \textbf{805}, 60 (2015).
\newblock \doi{10.1088/0004-637X/805/1/60}

\bibitem{Berkhuijsen_etal_1971}
E.M. {Berkhuijsen}, C.G.T. {Haslam}, C.J. {Salter}, \aap, \textbf{14}, 252
  (1971)

\bibitem{RogersonYork:1973ions}
J.B. {Rogerson}, D.G. {York}, J.F. {Drake}, E.B. {Jenkins}, D.C. {Morton},
  L.~{Spitzer}, \apjl, \textbf{181}, L110 (1973).
\newblock \doi{10.1086/181196}

\bibitem{RLIIItemp}
S.~{Redfield}, J.L. {Linsky}, \apj, \textbf{613}, 1004 (2004)

\bibitem{Frisch:2011araa}
P.C. {Frisch}, S.~Redfield, J.~Slavin, \araa, \textbf{49} (2011)

\bibitem{SlavinFrisch:2008}
J.D. {Slavin}, P.C. {Frisch}, \aap, \textbf{491}, 53 (2008)

\bibitem{Schwadron:2011sep}
N.A. {Schwadron}, F.~{Allegrini}, M.~{Bzowski}, E.R. {Christian}, G.B. {Crew},
  M.~{Dayeh}, R.~{DeMajistre}, P.~{Frisch}, H.O. {Funsten}, S.A. {Fuselier},
  K.~{Goodrich}, M.~{Gruntman}, P.~{Janzen}, H.~{Kucharek}, G.~{Livadiotis},
  D.J. {McComas}, E.~{Moebius}, C.~{Prested}, D.~{Reisenfeld}, M.~{Reno},
  E.~{Roelof}, J.~{Siegel}, R.~{Vanderspek}, \apj, \textbf{731}, 56 (2011).
\newblock \doi{10.1088/0004-637X/731/1/56}

\bibitem{MeyerPeek:2012leo}
D.M. {Meyer}, J.T. {Lauroesch}, J.E.G. {Peek}, C.~{Heiles}, \apj, \textbf{752},
  119 (2012).
\newblock \doi{10.1088/0004-637X/752/2/119}

\bibitem{Peeketal:2011}
J.E.G. {Peek}, C.~{Heiles}, K.M.G. {Peek}, D.M. {Meyer}, J.T. {Lauroesch}, \apj,
  \textbf{735}, 129 (2011).
\newblock \doi{10.1088/0004-637X/735/2/129}

\bibitem{Salvati:2010}
M.~{Salvati}, \aap, \textbf{513}, A28+ (2010)

\bibitem{McComas:2009sci}
D.J. {McComas}, F.~{Allegrini}, P.~{Bochsler}, M.~{Bzowski}, E.R. {Christian},
  G.B. {Crew}, R.~{DeMajistre}, H.~{Fahr}, H.~{Fichtner}, P.C. {Frisch}, H.O.
  {Funsten}, S.A. {Fuselier}, G.~{Gloeckler}, M.~{Gruntman}, J.~{Heerikhuisen},
  V.~{Izmodenov}, P.~{Janzen}, P.~{Knappenberger}, S.~{Krimigis},
  H.~{Kucharek}, M.~{Lee}, G.~{Livadiotis}, S.~{Livi}, R.J. {MacDowall},
  D.~{Mitchell}, E.~{M{\"o}bius}, T.~{Moore}, N.V. {Pogorelov},
  D.~{Reisenfeld}, E.~{Roelof}, L.~{Saul}, N.A. {Schwadron}, P.W. {Valek},
  R.~{Vanderspek}, P.~{Wurz}, G.P. {Zank}, Science, \textbf{326}, 959 (2009)

\bibitem{Schwadron:2009sci}
N.A. {Schwadron}, M.~{Bzowski}, G.B. {Crew}, M.~{Gruntman}, H.~{Fahr},
  H.~{Fichtner}, P.C. {Frisch}, H.O. {Funsten}, S.~{Fuselier},
  J.~{Heerikhuisen}, V.~{Izmodenov}, H.~{Kucharek}, M.~{Lee}, G.~{Livadiotis},
  D.J. {McComas}, E.~{Moebius}, T.~{Moore}, J.~{Mukherjee}, N.V. {Pogorelov},
  C.~{Prested}, D.~{Reisenfeld}, E.~{Roelof}, G.P. {Zank}, Science,
  \textbf{326}, 966 (2009)

\bibitem{Funsten:2013}
H.O. {Funsten}, R.~{DeMajistre}, P.C. {Frisch}, J.~{Heerikhuijsen}, D.M.
  {Higdon}, P.~{Janzen}, B.~{Larsen}, G.~{Livadiotis}, D.J. {McComas},
  E.~{M{\"o}bius}, C.~{Reese}, D.B. {Reisenfeld}, N.A. {Schwadron}, E.J.
  {Zirnstein}, \apj, \textbf{776}, 30 (2013)

\bibitem{Heerikhuisen_etal_2010}
J.~{Heerikhuisen}, N.V. {Pogorelov}, G.P. {Zank}, G.B. {Crew}, P.C. {Frisch},
  H.O. {Funsten}, P.H. {Janzen}, D.J. {McComas}, D.B. {Reisenfeld}, N.A.
  {Schwadron}, \apjl, \textbf{708}, L126 (2010)

\bibitem{Zirnstein:2016ismf}
E.J. {Zirnstein}, J.~{Heerikhuisen}, H.O. {Funsten}, G.~{Livadiotis}, D.J.
  {McComas}, N.V. {Pogorelov}, \apjl, \textbf{818}, L18 (2016).
\newblock \doi{10.3847/2041-8205/818/1/L18}

\bibitem{Frisch:inprep}
P.C. {Frisch}, A.~{Berdyugin}, V.~{Piirola}, A.M. {Magalhaes}, D.B.
  {Seriacopi}, S.J. {Wiktorowicz}, B.G. {Andersson}, H.O. {Funsten}, D.J.
  {McComas}, N.A. {Schwadron}, J.D. {Slavin}, A.J. {Hanson}, C.W. {Fu}, {In
  preparation}  (2016)

\bibitem{PerrotGrenier:2003}
C.A. {Perrot}, I.A. {Grenier}, \aap, \textbf{404}, 519 (2003).
\newblock \doi{10.1051/0004-6361:20030477}

\bibitem{Grenier:2004rev}
I.A. {Grenier}, ArXiv Astrophysics e-prints  (2004)

\bibitem{Bobylev:2014gbrev}
V.V. {Bobylev}, Astrophysics, \textbf{57}, 583 (2014).
\newblock \doi{10.1007/s10511-014-9360-7}

\bibitem{Alfaro:1991}
E.J. {Alfaro}, J.~{Cabrera-Cano}, A.J. {Delgado}, \apj, \textbf{378}, 106
  (1991).
\newblock \doi{10.1086/170410}

\bibitem{StothersFrogel:1974}
R.~{Stothers}, J.A. {Frogel}, \aj, \textbf{79}, 456 (1974)

\bibitem{Bobylev:2014gbvdbeltorionarm}
V.V. {Bobylev}, A.T. {Bajkova}, Astronomy Letters, \textbf{40}, 783 (2014).
\newblock \doi{10.1134/S1063773714120020}

\bibitem{Alfaro_etal_2009}
E.J. {Alfaro}, F.~{Elias}, J.~{Cabrera-Ca{\~n}o}, \apss, \textbf{324}, 141
  (2009).
\newblock \doi{10.1007/s10509-009-0119-2}

\bibitem{Elias_etal_2009}
F.~{Elias}, E.J. {Alfaro}, J.~{Cabrera-Ca{\~n}o}, \mnras, \textbf{397}, 2
  (2009).
\newblock \doi{10.1111/j.1365-2966.2009.14465.x}

\bibitem{Heiles:1979}
C.~{Heiles}, \apj, \textbf{229}, 533 (1979)

\bibitem{Heiles:1984worm}
C.~{Heiles}, \apjs, \textbf{55}, 585 (1984)

\bibitem{cox05}
D.P. {Cox}, \araa, \textbf{43}, 337 (2005).
\newblock \doi{10.1146/annurev.astro.43.072103.150615}

\bibitem{Heiles:1989shellbeta}
C.~{Heiles}, \apj, \textbf{336}, 808 (1989)

\bibitem{castoretal75}
J.~{Castor}, R.~{McCray}, R.~{Weaver}, \apjl, \textbf{200}, L107 (1975).
\newblock \doi{10.1086/181908}

\bibitem{weaveretal77}
R.~{Weaver}, R.~{McCray}, J.~{Castor}, P.~{Shapiro}, R.~{Moore}, \apj,
  \textbf{218}, 377 (1977).
\newblock \doi{10.1086/155692}

\bibitem{mvl84}
C.F. {McKee}, D.~{van Buren}, B.~{Lazareff}, \apjl, \textbf{278}, L115 (1984).
\newblock \doi{10.1086/184237}

\bibitem{km92a}
B.C. {Koo}, C.F. {McKee}, \apj, \textbf{388}, 93 (1992).
\newblock \doi{10.1086/171132}

\bibitem{km92b}
B.C. {Koo}, C.F. {McKee}, \apj, \textbf{388}, 103 (1992).
\newblock \doi{10.1086/171133}

\bibitem{vanmarleetal15}
A.J. {van Marle}, Z.~{Meliani}, A.~{Marcowith}, \aap, \textbf{584}, A49 (2015).
\newblock \doi{10.1051/0004-6361/201425230}

\bibitem{gml96a}
G.~{Garcia-Segura}, M.M. {Mac Low}, N.~{Langer}, \aap, \textbf{305}, 229 (1996)

\bibitem{glm96b}
G.~{Garcia-Segura}, N.~{Langer}, M.M. {Mac Low}, \aap, \textbf{316}, 133 (1996)

\bibitem{fhy03}
T.~{Freyer}, G.~{Hensler}, H.W. {Yorke}, \apj, \textbf{594}, 888 (2003).
\newblock \doi{10.1086/376937}

\bibitem{fhy06}
T.~{Freyer}, G.~{Hensler}, H.W. {Yorke}, \apj, \textbf{638}, 262 (2006).
\newblock \doi{10.1086/498734}

\bibitem{vvd07}
V.V. {Dwarkadas}, \apj, \textbf{667}, 226 (2007).
\newblock \doi{10.1086/520670}

\bibitem{tbfr90}
G.~{Tenorio-Tagle}, P.~{Bodenheimer}, J.~{Franco}, M.~{Rozyczka}, \mnras
  \textbf{244}, 563 (1990)

\bibitem{trfb91}
G.~{Tenorio-Tagle}, M.~{Rozyczka}, J.~{Franco}, P.~{Bodenheimer}, \mnras,
  \textbf{251}, 318 (1991).
\newblock \doi{10.1093/mnras/251.2.318}

\bibitem{rtfb93}
M.~{Rozyczka}, G.~{Tenorio-Tagle}, J.~{Franco}, P.~{Bodenheimer}, \mnras
  \textbf{261}, 674 (1993).
\newblock \doi{10.1093/mnras/261.3.674}

\bibitem{vvd05}
V.V. {Dwarkadas}, \apj, \textbf{630}, 892 (2005).
\newblock \doi{10.1086/432109}

\bibitem{vanmarleetal10}
A.J. {van Marle}, N.~{Smith}, S.P. {Owocki}, B.~{van Veelen}, \mnras,
  \textbf{407}, 2305 (2010).
\newblock \doi{10.1111/j.1365-2966.2010.16851.x}

\bibitem{cgg03}
Y.H. {Chu}, R.A. {Gruendl}, M.A. {Guerrero}, in \emph{Revista Mexicana de
  Astronomia y Astrofisica Conference Series}, \emph{Revista Mexicana de
  Astronomia y Astrofisica, vol.~27}, vol.~15, ed. by J.~{Arthur}, W.J.
  {Henney} (2003), \emph{Revista Mexicana de Astronomia y Astrofisica,
  vol.~27}, vol.~15, pp. 62--67

\bibitem{chuetal03}
Y.H. {Chu}, M.A. {Guerrero}, R.A. {Gruendl}, G.~{Garc{\'{\i}}a-Segura}, H.J.
  {Wendker}, \apj, \textbf{599}, 1189 (2003).
\newblock \doi{10.1086/379607}

\bibitem{toalaetal15}
J.A. {Toal{\'a}}, M.A. {Guerrero}, Y.H. {Chu}, R.A. {Gruendl}, \mnras,
  \textbf{446}, 1083 (2015).
\newblock \doi{10.1093/mnras/stu2163}

\bibitem{toalaetal16}
J.A. {Toal{\'a}}, M.A. {Guerrero}, Y.H. {Chu}, S.J. {Arthur}, D.~{Tafoya}, R.A.
  {Gruendl}, \mnras, \textbf{456}, 4305 (2016).
\newblock \doi{10.1093/mnras/stv2819}

\bibitem{ta11}
J.A. {Toal{\'a}}, S.J. {Arthur}, \apj, \textbf{737}, 100 (2011).
\newblock \doi{10.1088/0004-637X/737/2/100}

\bibitem{dr13}
V.V. {Dwarkadas}, D.L. {Rosenberg}, High Energy Density Physics \textbf{9}, 226
  (2013).
\newblock \doi{10.1016/j.hedp.2012.12.003}

\bibitem{dr15}
V.V. {Dwarkadas}, D.~{Rosenberg}, in \emph{Wolf-Rayet Stars: Proceedings of an
  International Workshop held in Potsdam, Germany, 1-5 June 2015. Edited by
  Wolf-Rainer Hamann, Andreas Sander, Helge Todt. Universit{\"a}tsverlag
  Potsdam, 2015., p.329-332}, ed. by W.R. {Hamann}, A.~{Sander}, H.~{Todt}
  (2015), pp. 329--332

\bibitem{mm88}
M.M. {Mac Low}, R.~{McCray}, \apj, \textbf{324}, 776 (1988).
\newblock \doi{10.1086/165936}

\bibitem{mm89}
M.M. {Mac Low}, R.~{McCray}, M.L. {Norman}, \apj, \textbf{337}, 141 (1989).
\newblock \doi{10.1086/167094}

\bibitem{yadavetal16}
N.~{Yadav}, D.~{Mukherjee}, P.~{Sharma}, B.B. {Nath}, ArXiv e-prints,  (2016)

\bibitem{korpietal99}
M.J. {Korpi}, A.~{Brandenburg}, A.~{Shukurov}, I.~{Tuominen}, \aap,
  \textbf{350}, 230 (1999)

\bibitem{shull93}
J.M. {Shull}, in \emph{Massive Stars: Their Lives in the Interstellar Medium},
  \emph{Astronomical Society of the Pacific Conference Series}, vol.~35, ed. by
  J.P. {Cassinelli}, E.B. {Churchwell} (1993), \emph{Astronomical Society of
  the Pacific Conference Series}, vol.~35, p. 327

\bibitem{og04}
M.S. {Oey}, G.~{Garc{\'{\i}}a-Segura}, \apj, \textbf{613}, 302 (2004).
\newblock \doi{10.1086/421483}

\bibitem{FerriereMacLow:1991}
K.M. {Ferriere}, M.~{Mac Low}, E.G. {Zweibel}, \apj, \textbf{375}, 239 (1991)

\bibitem{HanayamaTomiska:2006}
H.~{Hanayama}, K.~{Tomisaka}, \apj, \textbf{641}, 905 (2006).
\newblock \doi{10.1086/500527}

\bibitem{maclow99}
M.M. {Mac Low}, in \emph{New Perspectives on the Interstellar Medium},
  \emph{Astronomical Society of the Pacific Conference Series}, vol. 168, ed.
  by A.R. {Taylor}, T.L. {Landecker}, G.~{Joncas} (1999), \emph{Astronomical
  Society of the Pacific Conference Series}, vol. 168, p. 303

\bibitem{BrownDaviesHazard:1960}
R.~{Hanbury Brown}, R.D. {Davies}, C.~{Hazard}, The Observatory, \textbf{80},
  191 (1960)

\bibitem{Davies:1964}
R.D. {Davies}, \mnras, \textbf{128}, 173 (1964).
\newblock \doi{10.1093/mnras/128.2.173}

\bibitem{Berkhuijsen:1971}
E.M. {Berkhuijsen}, \aap, \textbf{14}, 359 (1971)

\bibitem{Salter:1983}
C.J. {Salter}, Bulletin of the Astronomical Society of India, \textbf{11}, 1
  (1983)

\bibitem{Heiles_etal_1980nps}
C.~{Heiles}, Y.. {Chu}, T.H. {Troland}, R.J. {Reynolds}, I.~{Yegingil}, \apj,
  \textbf{242}, 533 (1980)

\bibitem{Kun:2007loopIII}
M.~{Kun}, in \emph{Triggered Star Formation in a Turbulent ISM}, \emph{IAU
  Symposium}, vol. 237, ed. by B.G. {Elmegreen}, J.~{Palous} (2007), \emph{IAU
  Symposium}, vol. 237, pp. 119--123.
\newblock \doi{10.1017/S1743921307001329}

\bibitem{Reynolds:1984alpvir}
R.J. {Reynolds}, in \emph{NASA Conference Publication}, \emph{NASA Conference
  Publication}, vol. 2345, ed. by Y.~{Kondo}, F.C. {Bruhweiler}, B.D. {Savage}
  (1984), \emph{NASA Conference Publication}, vol. 2345

\bibitem{VidalDickinsonDaviesLeahy:2015}
M.~{Vidal}, C.~{Dickinson}, R.D. {Davies}, J.P. {Leahy}, \mnras, \textbf{452},
  656 (2015).
\newblock \doi{10.1093/mnras/stv1328}

\bibitem{Wolleben:2007}
M.~{Wolleben}, \apj, \textbf{664}, 349 (2007).
\newblock \doi{10.1086/518711}

\bibitem{Santos:2011}
F.P. {Santos}, W.~{Corradi}, W.~{Reis}, \apj, \textbf{728}, 104 (2011)

\bibitem{Egger:1995}
R.J. {Egger}, B.~{Aschenbach}, \aap, \textbf{294}, L25 (1995)

\bibitem{HeilesReachKoo:1996}
C.~{Heiles}, W.T. {Reach}, B.C. {Koo}, \apj, \textbf{466}, 191 (1996)

\bibitem{deGeus:1991}
E.J. {de Geus}, in \emph{The Formation and Evolution of Star Clusters},
  \emph{Astronomical Society of the Pacific Conference Series}, vol.~13, ed. by
  {K.~Janes} (1991), \emph{Astronomical Society of the Pacific Conference
  Series}, vol.~13, pp. 40--54

\bibitem{Crawford:1991}
I.A. {Crawford}, \aap, \textbf{247}, 183 (1991)

\bibitem{Blaauw:1964araa}
A.~{Blaauw}, \araa, \textbf{2}, 213 (1964).
\newblock \doi{10.1146/annurev.aa.02.090164.001241}

\bibitem{deGeus:1992}
E.J. {de Geus}, \aap, \textbf{262}, 258 (1992)

\bibitem{Pecaut2012}
M J. {Pecaut}, E E. {Mamajek}, E. J. {Bubar}, \apj,  \textbf{746}, 154 (2012)

\bibitem{Pecaut2016} 
M J. {Pecaut}, E E. {Mamajek}, \mnras,  \textbf{461}, 794 (2016)

\bibitem{Maiz-Apellaniz_2001}
J.~{Ma{\'{\i}}z-Apell{\'a}niz}, \apjl, \textbf{560}, L83 (2001)

\bibitem{frisch81}
P.C. {Frisch}, \nat, \textbf{293}, 377 (1981).
\newblock \doi{10.1038/293377a0}

\bibitem{Iwan:1980npsXrayloopI}
D..C. {Iwan}, \apj, \textbf{239}, 316 (1980)

\bibitem{Koutroumpa:2009lowkeV}
D.~{Koutroumpa}, R.~{Lallement}, J.C. {Raymond}, V.~{Kharchenko}, \apj,
  \textbf{696}, 1517 (2009).
\newblock \doi{10.1088/0004-637X/696/2/1517}

\bibitem{Sofue:2015nps}
Y.~{Sofue}, \mnras, \textbf{447}, 3824 (2015).
\newblock \doi{10.1093/mnras/stu2661}

\bibitem{Lallement:2016nps}
R.~{Lallement}, S.~{Snowden}, K.D. {Kuntz}, T.M. {Dame}, D.~{Koutroumpa},
  I.~{Grenier}, J.M. {Casandjian}, ArXiv e-prints,  (2016)

\bibitem{SuFinkbeiner:2010}
M.~{Su}, T.R. {Slatyer}, D.P. {Finkbeiner}, \apj, \textbf{724}, 1044 (2010)

\bibitem{BerdyuginPiirola:2014s1}
A.~{Berdyugin}, V.~{Piirola}, P.~{Teerikorpi}, \aap, \textbf{561}, A24 (2014).
\newblock \doi{10.1051/0004-6361/201322604}

\bibitem{Sun:2015faratomnps}
X.H. {Sun}, T.L. {Landecker}, B.M. {Gaensler}, E.~{Carretti}, W.~{Reich}, J.P.
  {Leahy}, N.M. {McClure-Griffiths}, R.M. {Crocker}, M.~{Wolleben},
  M.~{Haverkorn}, K.A. {Douglas}, A.D. {Gray}, \apj, \textbf{811}, 40 (2015).
\newblock \doi{10.1088/0004-637X/811/1/40}

\bibitem{Bowyeretal:1968}
C.S. {Bowyer}, G.B. {Field}, J.F. {Mack}, \nat, \textbf{217}, 32 (1968)

\bibitem{McCammon:1983}
D.~{McCammon}, D.N. {Burrows}, W.T. {Sanders}, W.L. {Kraushaar}, \apj,
  \textbf{269}, 107 (1983)

\bibitem{williamsonetal74}
F.O. {Williamson}, W.T. {Sanders}, W.L. {Kraushaar}, D.~{McCammon},
  R.~{Borken}, A.N. {Bunner}, \apjl, \textbf{193}, L133 (1974).
\newblock \doi{10.1086/181649}

\bibitem{sandersetal77}
W.T. {Sanders}, W.L. {Kraushaar}, J.A. {Nousek}, P.M. {Fried}, \apjl,
  \textbf{217}, L87 (1977).
\newblock \doi{10.1086/182545}

\bibitem{cr87}
D.P. {Cox}, R.J. {Reynolds}, \araa, \textbf{25}, 303 (1987).
\newblock \doi{10.1146/annurev.aa.25.090187.001511}

\bibitem{ms90}
D.~{McCammon}, W.T. {Sanders}, \araa, \textbf{28}, 657 (1990).
\newblock \doi{10.1146/annurev.aa.28.090190.003301}

\bibitem{Gehrels:1993}
N.~{Gehrels}, C.M. {Laird}, C.H. {Jackman}, J.K. {Cannizzo}, B.J. {Mattson},
  W.~{Chen}, \apj, \textbf{585}, 1169 (2003)

\bibitem{Frisch:1993}
P.C. {Frisch}, {Nature}, \textbf{364}, 395 (1993)

\bibitem{SmithCunha:1994}
V.V. {Smith}, K.~{Cunha}, B.~{Plez}, \aap, \textbf{281}, L41 (1994)

\bibitem{Pellizza:2005}
L.J. {Pellizza}, R.P. {Mignani}, I.A. {Grenier}, I.F. @~{Mirabel}, \aap,
  \textbf{435}, 625 (2005)

\bibitem{cs74}
D.P. {Cox}, B.W. {Smith}, \apjl, \textbf{189}, L105 (1974).
\newblock \doi{10.1086/181476}

\bibitem{Tetzlaffetal:2013}
N.~{Tetzlaff}, G.~{Torres}, R.~{Neuh{\"a}user}, M.M. {Hohle}, \mnras,
  \textbf{435}, 879 (2013).
\newblock \doi{10.1093/mnras/stt1358}

\bibitem{sc01}
R.K. {Smith}, D.P. {Cox}, \apjs, \textbf{134}, 283 (2001).
\newblock \doi{10.1086/320850}

\bibitem{ws09}
B.Y. {Welsh}, R.L. {Shelton}, \apss, \textbf{323}, 1 (2009).
\newblock \doi{10.1007/s10509-009-0053-3}

\bibitem{fuchsetal06}
B.~{Fuchs}, D.~{Breitschwerdt}, M.A. {de Avillez}, C.~{Dettbarn}, C.~{Flynn},
  \mnras, \textbf{373}, 993 (2006).
\newblock \doi{10.1111/j.1365-2966.2006.11044.x}

\bibitem{dab12}
M.A. {de Avillez}, D.~{Breitschwerdt}, \aap, \textbf{539}, L1 (2012).
\newblock \doi{10.1051/0004-6361/201117172}

\bibitem{Cravens_etal_2001}
T.E. {Cravens}, I.P. {Robertson}, S.L. {Snowden}, \jgr, \textbf{106}, 24883
  (2001)

\bibitem{KuntzSnowden:2015}
K.D. {Kuntz}, Y.M. {Collado-Vega}, M.R. {Collier}, H.K. {Connor}, T.E.
  {Cravens}, D.~{Koutroumpa}, F.S. {Porter}, I.P. {Robertson}, D.G. {Sibeck},
  S.L. {Snowden}, N.E. {Thomas}, B.M. {Walsh}, \apj, \textbf{808}, 143 (2015).
\newblock \doi{10.1088/0004-637X/808/2/143}

\bibitem{Frisch:2009ibex}
P.C. {Frisch}, M.~Bzowski, E.~{Gr{\"u}n}, V.~Izmodenov, H.~{Kr{\"u}ger}, J.L.
  Linsky, D.J. McComas, E.~M\"obius, S.~Redfield, N.~Schwadron, R.R. Shelton,
  J.D. Slavin, B.E. Wood, \ssr, \textbf{146}, 235 (2009)

\bibitem{SmithFosterBrickhouse:2014}
R.K. {Smith}, A.R. {Foster}, R.J. {Edgar}, N.S. {Brickhouse}, \apj,
  \textbf{787}, 77 (2014).
\newblock \doi{10.1088/0004-637X/787/1/77}

\bibitem{snowden15a}
S.L. {Snowden}, C.~{Heiles}, D.~{Koutroumpa}, K.D. {Kuntz}, R.~{Lallement},
  D.~{McCammon}, J.E.G. {Peek}, \apj, \textbf{806}, 119 (2015).
\newblock \doi{10.1088/0004-637X/806/1/119}

\bibitem{snowden15b}
S.L. {Snowden}, D.~{Koutroumpa}, K.D. {Kuntz}, R.~{Lallement},
  L.~{Puspitarini}, \apj, \textbf{806}, 120 (2015).
\newblock \doi{10.1088/0004-637X/806/1/120}

\bibitem{nehmeetal08a}
C.~{Nehm{\'e}}, C.~{Gry}, F.~{Boulanger}, J.~{Le Bourlot}, G.~{Pineau Des
  For{\^e}ts}, E.~{Falgarone}, \aap, \textbf{483}, 471 (2008).
\newblock \doi{10.1051/0004-6361:20078373}

\bibitem{gryetal09}
C.~{Gry}, C.~{Nehm{\'e}}, F.~{Boulanger}, J.~{Lebourlot}, G.P. {Des
  For{\^e}ts}, E.~{Falgarone}, in \emph{American Institute of Physics
  Conference Series}, \emph{American Institute of Physics Conference Series},
  vol. 1156, ed. by R.K. {Smith}, S.L. {Snowden}, K.D. {Kuntz} (2009),
  \emph{American Institute of Physics Conference Series}, vol. 1156, pp.
  218--222.
\newblock \doi{10.1063/1.3211817}

\bibitem{CoxHelenius:2003}
D.P. {Cox}, L.~{Helenius}, \apj, \textbf{583}, 205 (2003)

\bibitem{Spoelstra:1972lb}
T.A.T. {Spoelstra}, \aap, \textbf{21}, 61 (1972)

\bibitem{Berkhuijsen:1973r-I}
E.M. {Berkhuijsen}, \aap, \textbf{24}, 143 (1973)

\bibitem{Frisch:2010s1}
P.C. {Frisch}, \apj, \textbf{714}, 1679 (2010).
\newblock \doi{10.1088/0004-637X/714/2/1679}

\bibitem{Heiles:1998whence}
C.~{Heiles}, \apj, \textbf{498}, 689 (1998)

\bibitem{FrischSchwadron:2014icns}
P.C. {Frisch}, N.A. {Schwadron}, in \emph{{Outstanding Problems in
  Heliophysics: From Coronal Heating to the Edge of the Heliosphere}},
  \emph{Astronomical Society of the Pacific Conference Series}, vol. 484, ed.
  by Q.~{Hu}, G.P. {Zank} (2014), \emph{Astronomical Society of the Pacific
  Conference Series}, vol. 484, p.~42

\bibitem{RLIV:2008}
S.~{Redfield}, J.L. {Linsky}, \apj, \textbf{673}, 283 (2008)

\bibitem{Crutcher:1982}
R.M. {Crutcher}, \apj, \textbf{254}, 82 (1982)

\bibitem{FrischYork:1986}
P.~{Frisch}, D.G. {York}, in \emph{The {G}alaxy and the {S}olar {S}ystem},
  (University of Arizona Press, 1986), pp. 83--100

\bibitem{Bzowski:1988}
M.~{Bzowski}, Acta Astronomica, \textbf{38}, 443 (1988)

\bibitem{FGW:2002}
P.C. {Frisch}, L.~{Grodnicki}, D.E. {Welty}, \apj, \textbf{574}, 834 (2002)

\bibitem{GryJenkins:2014clic}
C.~{Gry}, E.B. {Jenkins}, \aap, \textbf{567}, A58 (2014).
\newblock \doi{10.1051/0004-6361/201323342}

\bibitem{Herbig:1968}
G.H. {Herbig}, Zeitschrift Astrophysics, \textbf{68}, 243 (1968)

\bibitem{Lallement:1986}
R.~{Lallement}, A.~{Vidal-Madjar}, R.~{Ferlet}, \aap, \textbf{168}, 225 (1986)

\bibitem{LallementBertin:1992}
R.~{Lallement}, P.~{Bertin}, \aap, \textbf{266}, 479 (1992)

\bibitem{Frisch:2003apex}
P.C. {Frisch}, \apj, \textbf{593}, 868 (2003)

\bibitem{Moebius:2004he}
M.A. {Kubiak}, P.~{Swaczyna}, M.~{Bzowski}, J.M. {Sok{\'o}{\l}}, S.A.
  {Fuselier}, A.~{Galli}, D.~{Heirtzler}, H.~{Kucharek}, T.W. {Leonard}, D.J.
  {McComas}, E.~{M{\"o}bius}, J.~{Park}, N.A. {Schwadron}, P.~{Wurz}, \apjs,
  \textbf{223}, 25 (2016).
\newblock \doi{10.3847/0067-0049/223/2/25}

\bibitem{Schwadron:2015He}
N.A. {Schwadron}, E.~{Moebius}, T.~{Leonard}, S.A. {Fuselier}, D.J. {McComas},
  D.~{Heirtzler}, H.~{Kucharek}, F.~{Rahmanifard}, M.~{Bzowski}, M.A. {Kubiak},
  J.~{Sokol}, P.~{Swaczyna}, P.~{Frisch}, \apjs, \textbf{220}, 25 (2015).
\newblock \doi{10.1088/0067-0049/220/2/25}

\bibitem{Frisch:1999}
P.C. {Frisch}, J.M. {Dorschner}, J.~{Geiss}, J.M. {Greenberg}, E.~{Gr\"un},
  M.~{Landgraf}, P.~{Hoppe}, A.P. {Jones}, W.~{Kr{\"{a}}tschmer}, T.J. {Linde},
  G.E. {Morfill}, W.~{Reach}, J.D. {Slavin}, J.~{Svestka}, A.N. {Witt}, G.P.
  {Zank}, \apj, \textbf{525}, 492 (1999)

\bibitem{KimuraMann:2003velorigin}
H.~{Kimura}, I.~{Mann}, E.K. {Jessberger}, \apj, \textbf{582}, 846 (2003)

\bibitem{Frisch:1979}
P.C. {Frisch}, \apj, \textbf{227}, 474 (1979)

\bibitem{Andersson:2015araa}
B.G. {Andersson}, A.~{Lazarian}, J.E. {Vaillancourt}, \araa, \textbf{53}, 501
  (2015).
\newblock \doi{10.1146/annurev-astro-082214-122414}

\bibitem{McComasLewisSchwadron:2014}
D.J. {McComas}, W.S. {Lewis}, N.A. {Schwadron}, Reviews of Geophysics,
  \textbf{52}, 118 (2014).
\newblock \doi{10.1002/2013RG000438}

\bibitem{SchwadronMcComas:2013ret}
N.A. {Schwadron}, D.J. {McComas}, \apj, \textbf{764}, 92 (2013).
\newblock \doi{10.1088/0004-637X/764/1/92}

\bibitem{Moebius:2009sci}
E.~{M{\"o}bius}, P.~{Bochsler}, M.~{Bzowski}, G.B. {Crew}, H.O. {Funsten}, S.A.
  {Fuselier}, A.~{Ghielmetti}, D.~{Heirtzler}, V.V. {Izmodenov}, M.~{Kubiak},
  H.~{Kucharek}, M.A. {Lee}, T.~{Leonard}, D.J. {McComas}, L.~{Petersen},
  L.~{Saul}, J.A. {Scheer}, N.~{Schwadron}, M.~{Witte}, P.~{Wurz}, Science
  \textbf{326}, 969 (2009).
\newblock \doi{10.1126/science.1180971}

\bibitem{Schwadron:2015H}
N.A. {Schwadron}, E.~{Moebius}, H.~{Kucharek}, M.A. {Lee}, J.~{French},
  L.~{Saul}, P.~{Wurz}, M.~{Bzowski}, S.A. {Fuselier}, G.~{Livadiotis}, D.J.
  {McComas}, P.~{Frisch}, M.~{Gruntman}, H.R. {Mueller}, \apj, \textbf{775}, 86
  (2013).
\newblock \doi{10.1088/0004-637X/775/2/86}

\bibitem{McComas:2015isn}
D.J. {McComas}, M.~{Bzowski}, P.~{Frisch}, S.A. {Fuselier}, M.A. {Kubiak},
  H.~{Kucharek}, T.~{Leonard}, E.~{M{\"o}bius}, N.A. {Schwadron}, J.M.
  {Sok{\'o}{\l}}, P.~{Swaczyna}, M.~{Witte}, \apj, \textbf{801}, 28 (2015).
\newblock \doi{10.1088/0004-637X/801/1/28}

\bibitem{Ebel:2000}
D.S. {Ebel}, \jgr, \textbf{105}, 10363 (2000)

\bibitem{RoutlySpitzer:1952}
P.~{Routly}, J.~{Spitzer}, L., \apj, \textbf{115}, 227 (1952)

\bibitem{Spitzer:1976}
L.~{Spitzer}, Comments on Astrophysics, \textbf{6}, 177 (1976)

\bibitem{SembachDanks:1994}
K.R. {Sembach}, A.C. {Danks}, \aap, \textbf{289}, 539 (1994)

\bibitem{SilukSilk:1974}
R.S. {Siluk}, J.~{Silk}, \apj, \textbf{192}, 51 (1974)

\bibitem{Jones:1996}
A.P. Jones, A.~Tielens, D.J. Hollenbach, \apj \textbf{469}, 740 (1996)

\bibitem{Slavin_etal_2015}
J.D. {Slavin}, E.~{Dwek}, A.P. {Jones}, \apj, \textbf{803}, 7 (2015).
\newblock \doi{10.1088/0004-637X/803/1/7}

\bibitem{Jenkins:2009}
E.B. {Jenkins}, \apj, \textbf{700}, 1299 (2009)

\bibitem{Welty:199923ori}
D.E. {Welty}, L.M. {Hobbs}, J.T. {Lauroesch}, D.C. {Morton}, L.~{Spitzer}, D.G.
  {York}, \apjs, \textbf{124}, 465 (1999)

\bibitem{McComas:2015six}
D.J. {McComas}, M.~{Bzowski}, S.A. {Fuselier}, P.C. {Frisch}, A.~{Galli}, V.V.
  {Izmodenov}, O.A. {Katushkina}, M.A. {Kubiak}, M.A. {Lee}, T.W. {Leonard},
  E.~{M{\"o}bius}, J.~{Park}, N.A. {Schwadron}, J.M. {Sok{\'o}{\l}},
  P.~{Swaczyna}, B.E. {Wood}, P.~{Wurz}, \apjs, \textbf{220}, 22 (2015).
\newblock \doi{10.1088/0067-0049/220/2/22}

\bibitem{Sterken:2015sixteen}
V.J. {Sterken}, P.~{Strub}, H.~{Kr{\"u}ger}, R.~{von Steiger}, P.~{Frisch},
  \apj, \textbf{812}, 141 (2015)

\bibitem{Altobelli:2016cassini}
N.~{Altobelli}, F.~{Postberg}, et~al., Science, \textbf{352}, 6283 (2016).
\newblock \doi{10.1126/science.aac6397}

\bibitem{Schwadron:2014sci}
N.A. {Schwadron}, F.C. {Adams}, E.R. {Christian}, P.~{Desiati}, P.~{Frisch},
  H.O. {Funsten}, J.R. {Jokipii}, D.J. {McComas}, E.~{Moebius}, G.P. {Zank},
  Science \textbf{343}, 988 (2014).
\newblock \doi{10.1126/science.1245026}

\bibitem{Fisk+Ramaty_1974}
L.A. Fisk, B.~Kozlovsky, R.~Ramaty, \apj, \textbf{190}, L35 (1974)

\bibitem{Gloeckler_Fisk_2007}
G.~{Gloeckler}, L.~{Fisk}, \ssr, \textbf{27}, 489 (2007)

\bibitem{Witte:2004}
M.~{Witte}, \aap, \textbf{426}, 835 (2004).
\newblock \doi{10.1051/0004-6361:20035956}

\bibitem{CummingsStone:2002}
A.C. {Cummings}, E.C. {Stone}, C.D. {Steenberg}, \apj, \textbf{578}, 194 (2002)

\bibitem{Frisch:1995rev}
P.C. {Frisch}, \ssr, \textbf{72}, 499 (1995)

\bibitem{Wolffetal:1999}
B.~{Wolff}, D.~{Koester}, R.~{Lallement}, \aap, \textbf{346}, 969 (1999)

\bibitem{Bzowski:2013neutralsurvival}
M.~{Bzowski}, J.M. {Sok{\'o}{\l}}, M.A. {Kubiak}, H.~{Kucharek}, \aap,
  \textbf{557}, A50 (2013).
\newblock \doi{10.1051/0004-6361/201321700}

\bibitem{Park:2014NeO}
J.~{Park}, H.~{Kucharek}, E.e.a. {M{\"o}bius}, \apj, \textbf{795}, 97 (2014)

\bibitem{Park:2015isn}
J.~{Park}, H.~{Kucharek}, E.~{M{\"o}bius}, A.~{Galli}, G.~{Livadiotis}, S.A.
  {Fuselier}, D.J. {McComas}, \apjs, \textbf{220}, 34 (2015).
\newblock \doi{10.1088/0067-0049/220/2/34}

\bibitem{Schwadron:2016O}
N.A. {Schwadron}, E.~{M{\"o}bius}, D.J. {McComas}, P.~{Bochsler}, M.~{Bzowski},
  S.A. {Fuselier}, G.~{Livadiotis}, P.~{Frisch}, H.R. {M{\"u}ller},
  D.~{Heirtzler}, H.~{Kucharek}, M.A. {Lee}, \apj, \textbf{828}, 81 (2016).
\newblock \doi{10.3847/0004-637X/828/2/81}

\bibitem{Slavin:1989}
J.D. {Slavin}, \apj, \textbf{346}, 718 (1989)

\bibitem{SlavinFrisch:2002}
J.D. {Slavin}, P.C. {Frisch}, \apj, \textbf{565}, 364 (2002)

\bibitem{Welty:2002zeta}
D.E. {Welty}, E.B. {Jenkins}, J.C. {Raymond}, C.~{Mallouris}, D.G. {York}, \apj,
  \textbf{579}, 304 (2002)

\bibitem{hhr99}
C.~{Heiles}, L.M. {Haffner}, R.J. {Reynolds}, in \emph{New Perspectives on the
  Interstellar Medium}, \emph{Astronomical Society of the Pacific Conference
  Series}, vol. 168, ed. by A.R. {Taylor}, T.L. {Landecker}, G.~{Joncas}
  (1999), \emph{Astronomical Society of the Pacific Conference Series}, vol.
  168, p. 211

\bibitem{obbt15}
B.B. {Ochsendorf}, A.G.A. {Brown}, J.~{Bally}, A.G.G.M. {Tielens}, \apj,
  \textbf{808}, 111 (2015).
\newblock \doi{10.1088/0004-637X/808/2/111}

\bibitem{ponetal16}
A.~{Pon}, B.B. {Ochsendorf}, J.~{Alves}, J.~{Bally}, S.~{Basu}, A.G.G.M.
  {Tielens}, \apj, \textbf{827}, 42 (2016).
\newblock \doi{10.3847/0004-637X/827/1/42}

\bibitem{rugeletal09}
G.~{Rugel}, T.~{Faestermann}, K.~{Knie}, G.~{Korschinek}, M.~{Poutivtsev},
  D.~{Schumann}, N.~{Kivel}, I.~{G{\"u}nther-Leopold}, R.~{Weinreich},
  M.~{Wohlmuther}, Physical Review Letters, \textbf{103}(7), 072502 (2009).
\newblock \doi{10.1103/PhysRevLett.103.072502}

\bibitem{lc03}
M.~{Limongi}, A.~{Chieffi}, \apj, \textbf{592}, 404 (2003).
\newblock \doi{10.1086/375703}

\bibitem{knieetal99}
K.~{Knie}, G.~{Korschinek}, T.~{Faestermann}, C.~{Wallner}, J.~{Scholten},
  W.~{Hillebrandt}, Physical Review Letters \textbf{83}, 18 (1999).
\newblock \doi{10.1103/PhysRevLett.83.18}

\bibitem{knieetal04}
K.~{Knie}, G.~{Korschinek}, T.~{Faestermann}, E.A. {Dorfi}, G.~{Rugel},
  A.~{Wallner}, Physical Review Letters, \textbf{93}(17), 171103 (2004).
\newblock \doi{10.1103/PhysRevLett.93.171103}

\bibitem{fhe05}
B.D. {Fields}, K.A. {Hochmuth}, J.~{Ellis}, \apj, \textbf{621}, 902 (2005).
\newblock \doi{10.1086/427797}

\bibitem{basuetal07}
S.~{Basu}, F.M. {Stuart}, C.~{Schnabel}, V.~{Klemm}, Physical Review Letters,
  \textbf{98}(14), 141103 (2007).
\newblock \doi{10.1103/PhysRevLett.98.141103}

\bibitem{fitoussietal08}
C.~{Fitoussi}, G.M. {Raisbeck}, K.~{Knie}, G.~{Korschinek}, T.~{Faestermann},
  S.~{Goriely}, D.~{Lunney}, M.~{Poutivtsev}, G.~{Rugel}, C.~{Waelbroeck},
  A.~{Wallner}, Physical Review Letters \textbf{101}(12), 121101 (2008).
\newblock \doi{10.1103/PhysRevLett.101.121101}

\bibitem{wallneretal16}
A.~{Wallner}, J.~{Feige}, N.~{Kinoshita}, M.~{Paul}, L.K. {Fifield},
  R.~{Golser}, M.~{Honda}, U.~{Linnemann}, H.~{Matsuzaki}, S.~{Merchel},
  G.~{Rugel}, S.G. {Tims}, P.~{Steier}, T.~{Yamagata}, S.R. {Winkler}, \nat,
  \textbf{532}, 69 (2016).
\newblock \doi{10.1038/nature17196}

\bibitem{bfsetal16}
D.~{Breitschwerdt}, J.~{Feige}, M.M. {Schulreich}, M.A.D. {Avillez},
  C.~{Dettbarn}, B.~{Fuchs}, \nat, \textbf{532}, 73 (2016).
\newblock \doi{10.1038/nature17424}

\bibitem{ffe16}
B.J. {Fry}, B.D. {Fields}, J.R. {Ellis}, \apj, \textbf{827}, 48 (2016).
\newblock \doi{10.3847/0004-637X/827/1/48}

\bibitem{fimianietal16}
L.~{Fimiani}, D.L. {Cook}, T.~{Faestermann}, J.M. {G{\'o}mez-Guzm{\'a}n},
  K.~{Hain}, G.~{Herzog}, K.~{Knie}, G.~{Korschinek}, P.~{Ludwig}, J.~{Park},
  R.C. {Reedy}, G.~{Rugel}, Physical Review Letters, \textbf{116}(15), 151104
  (2016).
\newblock \doi{10.1103/PhysRevLett.116.151104}

\bibitem{cooketal09}
D.L. {Cook}, E.~{Berger}, T.~{Faestermann}, G.F. {Herzog}, K.~{Knie},
  G.~{Korschinek}, M.~{Poutivtsev}, G.~{Rugel}, F.~{Serefiddin}, in \emph{Lunar
  and Planetary Science Conference}, \emph{Lunar and Planetary Inst.~Technical
  Report}, vol.~40 (2009), \emph{Lunar and Planetary Inst.~Technical Report},
  vol.~40, p. 1129

\bibitem{fimianietal12}
L.~{Fimiani}, D.L. {Cook}, T.~{Faestermann}, J.M. {Gomez Guzman}, K.~{Hain},
  G.F. {Herzog}, G.~{Korschinek}, B.~{Ligon}, P.~{Ludwig}, J.~{Park}, R.C.
  {Reedy}, G.~{Rugel}, in \emph{Lunar and Planetary Science Conference},
  \emph{Lunar and Planetary Inst.~Technical Report}, vol.~43 (2012),
  \emph{Lunar and Planetary Inst.~Technical Report}, vol.~43, p. 1279

\bibitem{blec13}
S.~{Bishop}, P.~{Ludwig}, R.~{Egli}, V.~{Chernenko}, T.~{Frederichs},
  S.~{Merchel}, G.~{Rugel}, in \emph{APS April Meeting Abstracts} (2013)

\bibitem{Ludwigetal16}
P.~{Ludwig}, S.~{Bishop}, R.~{Eglib}, V.e.a. {Chernenkoa}, PNAS, \textbf{113},
  9232 (2016).
\newblock \doi{10.1073/pnas.1601040113}

\bibitem{Kachelriessetal15}
M.~{Kachelriess}, A.~{Neronov}, and D.~V.~{Semikoz}, PRL, \textbf{115},
  181103 (2015)
\newblock \doi{10.1073/pnas.1601040113}


\bibitem{binnsetal16}
W.R. {Binns}, M.H. {Israel}, E.R. {Christian}, A.C. {Cummings}, G.A. {de
  Nolfo}, K.A. {Lave}, R.A. {Leske}, R.A. {Mewaldt}, E.C. {Stone}, T.T. {von
  Rosenvinge}, M.E. {Wiedenbeck}, Science, \textbf{352}, 677 (2016).
\newblock \doi{10.1126/science.aad6004}

\bibitem{Binns:2005}
W.R. {Binns}, M.E. {Wiedenbeck}, M.~{Arnould}, A.C. {Cummings}, J.S. {George},
  S.~{Goriely}, M.H. {Israel}, R.A. {Leske}, R.A. {Mewaldt}, G.~{Meynet}, L.M.
  {Scott}, E.C. {Stone}, T.T. {von Rosenvinge}, \apj, \textbf{634}, 351 (2005).
\newblock \doi{10.1086/496959}

\bibitem{Leske:2000acrisotope}
R.A. {Leske}, in \emph{26th International Cosmic Ray Conference, ICRC XXVI},
  \emph{American Institute of Physics Conference Series}, vol. 516, ed. by B.L.
  {Dingus}, D.B. {Kieda}, M.H. {Salamon} (2000), \emph{American Institute of
  Physics Conference Series}, vol. 516, pp. 274--282.
\newblock \doi{10.1063/1.1291481}

\bibitem{MurphyBinns:2016}
R.P. {Murphy}, M.~{Sasaki}, W.R. {Binns}, T.J. {Brandt}, T.~{Hams}, M.H.
  {Israel}, A.W. {Labrador}, J.T. {Link}, R.A. {Mewaldt}, J.W. {Mitchell}, B.F.
  {Rauch}, K.~{Sakai}, E.C. {Stone}, C.J. {Waddington}, N.E. {Walsh}, J.E.
  {Ward}, M.E. {Wiedenbeck}, ArXiv e-prints,  (2016)

\bibitem{wbc99}
M.E. {Wiedenbeck}, W.R. {Binns}, E.R. {Christian}, A.C. {Cummings}, B.L.
  {Dougherty}, P.L. {Hink}, J.~{Klarmann}, R.A. {Leske}, M.~{Lijowski}, R.A.
  {Mewaldt}, E.C. {Stone}, M.R. {Thayer}, T.T. {von Rosenvinge}, N.E.
  {Yanasak}, \apjl, \textbf{523}, L61 (1999).
\newblock \doi{10.1086/312242}

\bibitem{nm16}
A.~{Neronov}, G.~{Meynet}, \aap, \textbf{588}, A86 (2016).
\newblock \doi{10.1051/0004-6361/201527762}

\bibitem{Rauch:2009gcrdust}
B.F. {Rauch}, J.T. {Link}, K.~{Lodders}, M.H. {Israel}, L.M. {Barbier}, W.R.
  {Binns}, E.R. {Christian}, J.R. {Cummings}, G.A. {de Nolfo}, S.~{Geier}, R.A.
  {Mewaldt}, J.W. {Mitchell}, S.M. {Schindler}, L.M. {Scott}, E.C. {Stone},
  R.E. {Streitmatter}, C.J. {Waddington}, M.E. {Wiedenbeck}, \apj, \textbf{697},
  2083 (2009).
\newblock \doi{10.1088/0004-637X/697/2/2083}

\bibitem{epstein80}
R.I. {Epstein}, \mnras, \textbf{193}, 723 (1980).
\newblock \doi{10.1093/mnras/193.4.723}

\bibitem{bc81}
J.P. {Bibring}, C.J. {Cesarsky}, International Cosmic Ray Conference,
  \textbf{2}, 289 (1981)

\bibitem{EllisonDruryMeyer:1997}
D.C. {Ellison}, L.O. {Drury}, J.~{Meyer}, \apj, \textbf{487}, 197 (1997)

\bibitem{BarlowSilk:1977}
M.J. {Barlow}, J.~{Silk}, \apjl, \textbf{211}, L83 (1977).
\newblock \doi{10.1086/182346}

\bibitem{DraineSalpeter:1979a}
B.T. {Draine}, E.E. {Salpeter}, \apj, \textbf{231}, 438 (1979)

\bibitem{Jones:1994}
A.P. {Jones}, A.G.G.M. {Tielens}, D.J. {Hollenbach}, C.F. {McKee}, \apj
  \textbf{433}, 797 (1994)

\bibitem{knodlseder99}
J.~{Kn{\"o}dlseder}, \apj, \textbf{510}, 915 (1999).
\newblock \doi{10.1086/306601}

\bibitem{knodlsederetal99}
J.~{Kn{\"o}dlseder}, K.~{Bennett}, H.~{Bloemen}, R.~{Diehl}, W.~{Hermsen},
  U.~{Oberlack}, J.~{Ryan}, V.~{Sch{\"o}nfelder}, P.~{von Ballmoos}, \aap,
  \textbf{344}, 68 (1999)

\bibitem{diehletal10}
R.~{Diehl}, M.G. {Lang}, P.~{Martin}, H.~{Ohlendorf}, T.~{Preibisch},
  R.~{Voss}, P.~{Jean}, J.P. {Roques}, P.~{von Ballmoos}, W.~{Wang}, \aap,
  \textbf{522}, A51 (2010).
\newblock \doi{10.1051/0004-6361/201014302}

\bibitem{KrauseDiehl:2015}
M.G.H. {Krause}, R.~{Diehl}, Y.~{Bagetakos}, E.~{Brinks}, A.~{Burkert},
  O.~{Gerhard}, J.~{Greiner}, K.~{Kretschmer}, T.~{Siegert}, \aap, \textbf{578},
  A113 (2015).
\newblock \doi{10.1051/0004-6361/201525847}

\bibitem{HoyleLittleton:1939climate}
F.~{Hoyle}, R.A. {Lyttleton}, Proceedings of the Cambridge Philosophical
  Society \textbf{35}, 405 (1939).
\newblock \doi{10.1017/S0305004100021150}

\bibitem{McCrea:1977snrclimate}
D.H. {Clark}, W.H. {McCrea}, F.R. {Stephenson}, \nat, \textbf{265}, 318 (1977)

\bibitem{Frisch:1993gstar}
P.C. {Frisch}, \apj, \textbf{407}, 198 (1993)

\bibitem{Scherer:2006helioclimate}
K.~{Scherer}, H.~{Fichtner}, T.~{Borrmann}, J.~{Beer}, L.~{Desorgher},
  E.~{Fl{\"u}kiger}, H.J. {Fahr}, S.E.S. {Ferreira}, U.W. {Langner}, M.S.
  {Potgieter}, B.~{Heber}, J.~{Masarik}, N.~{Shaviv}, J.~{Veizer}, \ssr,
  \textbf{127}, 327 (2006).
\newblock \doi{10.1007/s11214-006-9126-6}

\bibitem{Frisch:2006book}
P.C. {Frisch}, \emph{{Solar Journey: The Significance of Our Galactic
  Environment for the Heliosphere and Earth, }} ({Springer}, 2006)

\bibitem{Holzer:1989}
T.E. {Holzer}, \araa, \textbf{27}, 199 (1989)

\bibitem{ZankFrisch:1999}
G.P. {Zank}, P.C. {Frisch}, \apj, \textbf{518}, 965 (1999)

\bibitem{Zank:2015araa}
G.P. {Zank}, \araa, \textbf{53}, 449 (2015).
\newblock \doi{10.1146/annurev-astro-082214-122254}

\bibitem{FrischMueller:2011ssr}
P.C. {Frisch}, H.R. {Mueller}, \ssr, pp. 130--+ (2011).
\newblock \doi{10.1007/s11214-011-9776-x}

\bibitem{WashimiTanaka:1999}
H.~{Washimi}, T.~{Tanaka}, Advances in Space Research, \textbf{23}, 551 (1999)

\bibitem{PogorelovSuess:2013solarcycle}
N.V. {Pogorelov}, S.T. {Suess}, S.N. {Borovikov}, R.W. {Ebert}, D.J. {McComas},
  G.P. {Zank}, \apj, \textbf{772}, 2 (2013).
\newblock \doi{10.1088/0004-637X/772/1/2}

\bibitem{Fields:2008snhelio}
B.D. {Fields}, T.~{Athanassiadou}, S.R. {Johnson}, \apj, \textbf{678}, 549-562
  (2008).
\newblock \doi{10.1086/523622}

\bibitem{Muelleretal:2008time}
H.R. {M\"uller}, P.C. {Frisch}, B.D. Fields, G.P. Zank, \ssr, p. 163 (2008)

\bibitem{SlavinFrisch:1996vela}
J.D. {Slavin}, P.C. {Frisch}, \ssr, \textbf{78}, 223 (1996)

\bibitem{thomas17}
B.~C.~Thomas, Astrobiology,  arXiv:1711.00410

\bibitem{FangJiang:2014_3C273}
T.~{Fang}, X.~{Jiang}, \apjl \textbf{785}, L24 (2014).
\newblock \doi{10.1088/2041-8205/785/2/L24}

\bibitem{Sembach:1997LoopI_IV}
K.R. {Sembach}, B.D. {Savage}, T.M. {Tripp}, \apj \textbf{480}, 216 (1997).
\newblock \doi{10.1086/303956}

\end{thebibliography}
\end{document}